\documentclass[a4paper,12pt,english]{article}

\usepackage{amsfonts,bm,amssymb,euscript,array,babel}
\usepackage[nosort]{cite}
\usepackage{mathtools}
\usepackage{tikz-cd}
\usepackage{amsmath}
\usepackage{hhline}
\usepackage[amsmath]{empheq}
\usepackage{hyperref}
\usepackage{cancel}
\usepackage{caption}
\usepackage{enumitem}
\usepackage{mathrsfs}
\usepackage{pdfsync}
\usepackage{tikz}

\newcolumntype{C}{>{$}c<{$}}

\usepackage{accents}

\newsavebox{\mysaveboxM}
\newsavebox{\mysaveboxT}

\makeatletter

\newcommand{\dd}{\mathrm{d}}

\newcommand{\w}{\wedge}
\newcommand{\bbm}{\left(\begin{matrix}}
\newcommand{\ebm}{\end{matrix}\right)}
\newcommand{\beq}{\begin{eqnarray}}
\newcommand{\eeq}{\end{eqnarray}}
\newcommand{\T}{\text{T}}

\DeclareMathOperator{\gh}{gh}

\makeatother

\newcommand{\sfrac}[2]{{\textstyle\frac{#1}{#2}}}

\newcommand{\be}{\begin{equation}}
\newcommand{\ee}{\end{equation}}

\newcommand{\beqa}{\begin{eqnarray}}
\newcommand{\eeqa}{\end{eqnarray}} 

\def\nn{\nonumber} \def \bea{\begin{eqnarray}} \def\eea{\end{eqnarray}}

\newcommand{\barr}{\begin{array}}
\newcommand{\earr}{\end{array}}
\numberwithin{equation}{section}

    \def\r{\rho}
\def\s{\sigma} \def\S{\Sigma}  

 \def\one{\mbox{1 \kern-.59em {\rm l}}}

\def\bit{\begin{itemize}} \def\eit{\end{itemize}}

\def\({\left(} \def\){\right)}

 \allowdisplaybreaks[3]

\numberwithin{equation}{section}

\begin{document}
	

\makeatother


\parindent=0cm

\renewcommand{\title}[1]{\vspace{10mm}\noindent{\Large{\bf

#1}}\vspace{8mm}} \newcommand{\authors}[1]{\noindent{\large

#1}\vspace{5mm}} \newcommand{\address}[1]{{\itshape #1\vspace{2mm}}}


\begin{titlepage}

\begin{flushright}
\small
RBI--ThPhys--2020--04
\end{flushright}

\begin{center}


\title{ {\Large 
{Courant sigma model and $L_\infty$-algebras}}}

\vskip 3mm

  \authors{ 
\large
    Clay James {Grewcoe}{\footnote{cgrewcoe@irb.hr}}
    , \ 
    Larisa Jonke{\footnote{larisa@irb.hr}} }
 
  \address{ Division of Theoretical Physics, Rudjer Bo\v skovi\'c Institute \\ Bijeni\v cka 54, 10000 Zagreb, Croatia \\
 
 }

\vskip 2cm

\begin{abstract}
\noindent
The Courant sigma model is a 3-dimensional topological sigma model of AKSZ type which has been used for the systematic description of closed strings in non-geometric flux backgrounds. In particular, the expression for the fluxes and their Bianchi identities coincide with the local form of the axioms of a Courant algebroid. On the other hand, the axioms of a Courant algebroid also coincide with the conditions for
gauge  invariance  of the Courant sigma model. 
In this paper we embed this interplay between background fluxes of closed strings, gauge (or more precisely  BRST) symmetries of the Courant sigma model and axioms of a Courant algebroid into an $L_\infty$-algebra structure. We show how the complete BV-BRST formulation of the Courant sigma model is described in terms of $L_\infty$-algebras. Moreover, the morphism between the $L_\infty$-algebra for a Courant algebroid and the one for the corresponding sigma model is constructed.

\end{abstract}

\end{center}

\vskip 2cm

\end{titlepage}

\setcounter{footnote}{0}
\tableofcontents


\section{Introduction}

Symmetries in general and gauge symmetries in particular are an indispensable tool for a theoretical description of a wide range of physical phenomena. Recent  research efforts are focused on a better understanding of  higher gauge symmetries and dualities in  field and string theories. A very general mathematical framework that encompasses these generalised concepts of symmetry is that of $L_\infty$-algebras.\footnote{The history of development of the concept  together with relevant references is
 given in \url{https://ncatlab.org/nlab/show/L-infinity-algebra\#History}} $L_\infty$-algebras are a generalisation of standard Lie algebras  in which the failure of the Jacobi identity for 2-brackets is controlled by higher 3-brackets, the failure of higher Jacobi identities for 3-brackets is controlled by 4-brackets and so on. The exact relations between higher brackets  defined on a graded vector space are given by the defining homotopy relations of the $L_\infty$-algebra.  Moreover, an $L_\infty$-algebra structure can be defined to include not only  symmetries of the theory, but also covariant equations of motion and conservation laws.  Provided that in  addition    a compatible  inner product exists, one can also write the action in terms of an $L_\infty$-algebra structure. 

 In physics,   $L_\infty$-algebras were identified as relevant algebraic structures in the construction of closed string field theory~\cite{csft,ls}. Already there it was realised that there is an intimate connection of $L_\infty$-algebras with the BV-BRST approach for quantisation of gauge theories. Recently, it was proposed  that $L_\infty$-algebras could  provide a classification of perturbative gauge theories in general~\cite{OB}. Furthermore, it was shown that using the $L_\infty$-algebra framework one can bootstrap consistent gauge theories~\cite{rb}. Choosing initial data of the theory in the form of 1- and 2-brackets one can  bootstrap higher brackets using the homotopy relations of an $L_\infty$-algebra and thus find consistent, gauge invariant theories defined by their equations of motion. This is reminiscent of the deformation of a free gauge theory into an interacting one in the BV-BRST approach. There one starts with a free, kinetic part of the action and its gauge  symmetry and adds systematically  all possible interaction terms consistent with BRST invariance, see review Ref.\cite{glenn} for more details and references.

Starting from closed string field theory, an attempt to make T-duality of string theory manifest  resulted in the construction of double field theory, which is a field theory for the massless sector of string theory defined on a doubled space~\cite{S1,S2,HZ1,HZ2}. The symmetry of this theory is generated by generalised diffeomorphisms which include standard diffeomorphisms and 2-form gauge transformations. It was shown that this algebra of symmetries is not of standard Lie-algebra type, but can be described by the structure of a
Courant algebroid. A Courant algebroid~\cite{C90,LWX,Sev} is defined on a generalised tangent bundle equipped with a 2-bracket which does not satisfy the Jacobi identity, a symmetric bilinear form and an anchor  map to the tangent bundle. It has been shown that its algebraic structure can be described as 2-term  $L_\infty$-algebra~\cite{Roytenberg:1998vn}. 

Furthermore, in Ref.\cite{Roytenberg:2006qz} Roytenberg  has shown that given the data of
a Courant algebroid one can uniquely construct the Batalin-Vilkovisky (BV) master action for a
membrane sigma model which is a first-order functional for generalised Wess-Zumino terms in
three dimensions.\footnote{See also Refs.~\cite{P1,Ikeda,HP}  for earlier related work.} This Courant
sigma model belongs to a general class of topological sigma models of AKSZ type~\cite{AKSZ} satisfying
the classical master equation. The membrane sigma models were subsequently used for
a systematic description of closed strings in non-geometric flux backgrounds~\cite{Hal, Mylonas, ChJL, Watamura, p12}. In particular, the expression for the fluxes and their Bianchi identities coincide with the local form of the axioms of a Courant algebroid. On the other hand, the axioms of a Courant algebroid also coincide with the conditions for
gauge  invariance and on-shell closure of the algebra of gauge transformations of the Courant sigma model. 

In this paper we would like to embed this intersection between background fluxes of closed strings, gauge (or BRST) symmetries of the Courant sigma model and axioms of a Courant algebroid into an $L_\infty$-algebra structure.\footnote{Similar ideas in the context of  topological open membranes where discussed in Ref.~\cite{HP}.} We shall explicitly construct the $L_\infty$-algebra structure for the classical, bosonic Courant sigma model  and then show how to extend this construction to include  the full BV-BRST action.  In the next section we shall introduce the necessary definitions and conventions closely following Ref.~\cite{brano}. In section 3 we shall construct the $L_\infty$-algebra for the classical Courant sigma model, including fields, symmetries and the action functional.\footnote{The general discussion of  $L_\infty$-algebra structures for AKSZ models (as QP-manifolds)  was already presented in the original paper ~\cite{AKSZ}.} Moreover, we shall demonstrate how homotopy identities naturally generate the axioms of a Courant algebroid in the form of coordinate expressions.  It is important to note that at the classical level we are discussing two different $L_\infty$-algebras, the $L_\infty$ gauge algebra $(\mathsf{L},\mu_i)$ and the tensor product algebra of $L_\infty$-algebra-valued de Rham forms $(\Omega^\bullet(M,\mathsf{L}),\mu'_i)$.  Next, we shall present the full BV-BRST action for the Courant sigma model reproducing Roytenbergs result~\cite{Roytenberg:2006qz} by explicitly constructing the $L_\infty$-algebra, this one including the complete BV-BRST complex. In the last section we present the mappings or  $L_\infty$-algebra morphism between the Courant algebroid $L_\infty$-algebra and the gauge algebra $(\mathsf{L},\mu_i)$ we constructed for the Courant sigma model. In order to construct these mappings we need to extend the $L_\infty$-algebra given in Ref.\cite{Roytenberg:1998vn} which essentially used a graded vector space concentrated in just two degrees, whereas the physical fields in  $(\mathsf{L},\mu_i)$  live in a graded vector space of three homogeneous subspaces. The construction of this morphism can be thought of as reproducing Roytenbergs result~\cite{Roytenberg:2006qz}  that given the data of a Courant algebroid one can uniquely construct the Courant sigma model (up to the additional structure of measure on the source space) now fully in the $L_\infty$-algebra formalism. Finally, in the appendices we state the relationship between different conventions and present some explicit calculations.

\section{On  $L_\infty$-algebras}

Once we include higher-degree gauge fields in our physical models, it is, in general, necessary to extend the standard description of symmetries based on Lie algebras and Lie groups to higher structures. Higher gauge fields appear naturally  in string theory, but one can easily embed this generalisation into the standard field-theoretical framework,  e.g. Refs.\cite{gs,glob}. A systematic  approach  to  these higher structures  can be formulated using $L_\infty$-algebras. In this section we shall briefly review the basic definitions and properties of  $L_\infty$-algebras  following the    conventions of \cite{brano}. The relation to other conventions, used for example  in Ref.\cite{OB}, is given in Appendix \ref{app:conventions}.

A ${L_\infty}${-algebra} or strong homotopy Lie algebra $(\mathsf{L},\mu_i)$ is a  graded vector space $\mathsf{L}$  with a collection of higher products that are graded totally antisymmetric  multilinear maps 
	\be
	\mu_i:\underbrace{\mathsf{L}\times\cdots\times \mathsf{L}}_{\text{$i$-times}}\to \mathsf{L}. \nonumber
	\ee
	of degree $2-i$ which satisfy the homotopy Jacobi identities:
	\begin{align}
	\sum_{j+k=i}\sum_\sigma\chi(\sigma;l_1,\ldots,l_i)(-1)^{k}\mu_{k+1}(\mu_j(l_{\sigma(1)},\ldots,l_{\sigma(j)}),l_{\sigma(j+1)},\ldots,l_{\sigma(i)})&=0;\label{eq:homotopyjac}\\
	\forall i\in\mathbb{N}\quad \forall l_1,\ldots,l_i&\in\mathsf{L}\nonumber.
	\end{align}
	Here $\chi(\sigma;l_1,\ldots,l_i)$ is the graded Koszul sign that includes the sign from the parity of the permutation of $\{1,\ldots,i\}$,  $\sigma$, ordered as: $\sigma(1)<\cdots<\sigma(j)$ and $\sigma(j+1)<\cdots<\sigma(i)$.\footnote{Permutations ordered in this way are conventionally called unshuffles.} By graded totally antisymmetric map we mean 
\begin{align*}
	\mu_i(\ldots,l_n,l_m,\ldots)&=(-1)^{|l_n||l_m|+1}\mu_i(\ldots,l_m,l_n,\ldots) ,
	\end{align*}
where $|l_n|$  is the degree of element $l_n\in \mathsf{L}$.

It is instructive to explicitly write the first three relations in \eqref{eq:homotopyjac}:
\begin{align*}
{ i=1:}\;\;\mu_1(\mu_1(l)) =0.\end{align*}
This relation states the map $\mu_1$ is a differential on $\mathsf{L}$.
\[
{ i=2:}\;\;
\mu_1(\mu_2(l_1,l_2))=\mu_2(\mu_1(l_1),l_2)-(-1)^{|l_1||l_2|}\mu_2(\mu_1(l_2),l_1).\]
From the second relation it is obvious the map $\mu_1$  is a derivation with respect to the graded 2-bracket $\mu_2$ on $\mathsf{L}$, while the third relation:
\begin{align*}
{ i=3:}\;\;\mu_1(\mu_3(l_1,l_2,l_3))&=\mu_2(\mu_2(l_1,l_2),l_3)-(-1)^{|l_2||l_3|}\mu_2(\mu_2(l_1,l_3),l_2)+{}\\
&\phantom{\,=\,}+(-1)^{|l_1|(|l_2|+|l_3|)}\mu_2(\mu_2(l_2,l_3),l_1)-\mu_3(\mu_1(l_1),l_2,l_3)+{}\\
&\phantom{\,=\,}+(-1)^{|l_1||l_2|}\mu_3(\mu_1(l_2),l_1,l_3)-(-1)^{|l_3|(|l_1|+|l_2|)}\mu_3(\mu_1(l_3),l_1,l_2),
\end{align*}
is the Jacobi identity for a 2-bracket $\mu_2$  up to homotopy given by $\mu_3$. Should the maps $\mu_i$ for $i\geqslant 3$ all be zero one would recover the standard differential graded Lie algebra.

An important class of ${L_\infty}${-algebras} that are  relevant for our construction are induced by  the tensor product of an $L_\infty$-algebra  with a differential graded commutative algebra.  In particular, if we take a de Rham complex on a manifold $M$, $(\Omega^\bullet(M),\dd)$   and tensor it with an ${L_\infty}${-algebra} we again obtain an $L_\infty$-algebra $(\mathsf{L'},\mu'_i)$:
\[
\mathsf{L'}\equiv\Omega^\bullet(M,\mathsf{L})\equiv\bigoplus_{k\in\mathbb{Z}}\Omega^\bullet_k(M,\mathsf{L}),\quad \Omega^\bullet_k(M,\mathsf{L})\equiv\bigoplus_{i+j=k}\Omega^i(M)\otimes\mathsf{L}_j,
\]
with higher products:
\begin{align}
\mu'_1(\alpha_1\otimes l_1)&=\dd \alpha_1\otimes l_1+(-1)^{|\alpha_1|}\alpha_1\otimes\mu_1(l_1),\label{eq:muprime1}\\
\mu'_i(\alpha_1\otimes l_1,\ldots,\alpha_i\otimes l_i)&=\!
\begin{aligned}[t]
(-1)^{i\sum_{j=1}^{i}|\alpha_j|+\sum_{j=0}^{i-2}|\alpha_{i-j}|\sum_{k=1}^{i-j-1}|l_k|}(\alpha_1\wedge\cdots \wedge\alpha_i)\otimes\mu_i(l_1,\ldots,l_i),&\label{eq:muprimei}\\
\forall i\geqslant 2,\quad \alpha_1,\ldots,\alpha_i\in\Omega^\bullet(M),\quad l_1,\ldots,l_i\in\mathsf{L}.&
\end{aligned}
\end{align}

Within the framework of ${L_\infty}${-algebras} one can define a generalisation of the Maurer-Cartan (MC) equation as follows. Take  $(\mathsf L', \mu'_i)$ and an element $a\in\mathsf{L'}_1$ which we call a gauge potential.  One defines the corresponding curvature as: 
	\be\label{eq:mceom}
	f\equiv\mu'_1(a)+\frac 1 2 \mu'_2(a,a)+\cdots=\sum_{i\geqslant1}\frac{1}{i!}\mu'_i(a,\ldots,a) .
	\ee
The generalised or homotopy Maurer-Cartan equation is then $f=0$.

 {Gauge transformations} of gauge potentials $a$ and their curvatures $f$ are given by:
\begin{align}
\delta_{c_0}a&=\sum_{i\geqslant0}\frac{1}{i!}\mu'_{i+1}(a,\ldots,a,c_0),\label{eq:gaugea}\\
\delta_{c_0}f&=\sum_{i\geqslant0}\frac{1}{i!}\mu'_{i+2}(a,\ldots,a,f,c_0),\nonumber
\end{align}
where $c_0\in\mathsf{L}_0$ is the \emph{level 0} gauge parameter. If a theory contains higher gauge symmetries we will have higher (\emph{level $k$}) gauge parameters $c_{-k}\in\mathsf{L}_{-k}$, $k>0$, with infinitesimal gauge transformations given by:
\[
\delta_{c_{-k-1}}c_{-k}=\sum_{i\geqslant0}\frac{1}{i!}\mu'_{i+1}(a,\ldots,a,c_{-k-1}).
\]

One can show that the algebra of gauge transformations closes up to terms proportional to the curvature $f$. In this respect, the MC equation can be interpreted either as an equation of motion or as a constraint on the kinematical data of the theory.  If we choose to think of the MC equation as  dynamical, the next question is if there exists an action from which this MC equation follows by variational principle. The answer is yes if it is possible to define a certain bilinear pairing compatible with the ${L_\infty}${-algebra} structure. In that case one defines a  cyclic  ${L_\infty}${-algebra} as an $L_\infty$-algebra (over $\mathbb{R}$) with a graded symmetric  non-degenerate bilinear pairing 
	\[
	\langle \,\cdot\,, \,\cdot\, \rangle_{\mathsf{L}}:\mathsf{L}\times\mathsf{L}\to\mathbb{R},\; \langle l_n,l_m\rangle_{\mathsf{L}}=(-1)^{|l_n||l_m|}\langle l_m,l_n\rangle_{\mathsf{L}} ,
	\]
	that satisfies the cyclicity condition:
	\begin{align}
	\langle l_1,\mu_i(l_2,\ldots,l_{i+1})\rangle_{\mathsf{L}}=(-1)^{i+i(|l_1|+|l_{n+1}|)|l_{i+1}|\sum_{j=1}^{i}|l_j|}\langle l_{i+1},\mu_i(l_1,\ldots,l_i)\rangle_{\mathsf{L}}&;\label{eq:cyclic}\\
	\forall i\in\mathbb{N}\nonumber&.
	\end{align}
If we have a tensored structure like $(\mathsf L', \mu'_i)$  then $(\mathsf L', \mu'_i)$  is cyclic provided  $(\mathsf{L},\mu)$ is cyclic and $M$ is an oriented, compact cycle. The induced inner product is then
\begin{align}
\langle\alpha_1\otimes l_1,\alpha_2\otimes l_2\rangle_{\mathsf{L'}}=(-1)^{|\alpha_2||l_1|}\int_M\alpha_1\wedge\alpha_2\;\langle l_1,l_2\rangle_{\mathsf{L}}.\label{eq:lprimepairing}
\end{align}
An inner product defined in this way allows us to write the action whose stationary point is the MC equation:
	\begin{align}
	S_{\mathrm{MC}}[a]\equiv\sum_{i\geqslant1}\frac{1}{(i+1)!}\langle a,\mu'_i(a,\ldots,a)\rangle_{\mathsf{L}'}.\label{eq:mcaction}
	\end{align}
It follows from the homotopy Jacobi identities \eqref{eq:homotopyjac} that this action is gauge invariant with respect to the variation of field $a$ as given in \eqref{eq:gaugea}.

\section{$L_\infty$ for the classical Courant sigma model}\label{sec:csm}

The Courant sigma model has been constructed in Ref.\cite{Ikeda} starting from Chern-Simons theory coupled to BF theory in BRST formalism, and in Ref.\cite{Roytenberg:2006qz} using the general construction for  AKSZ topological sigma models~\cite{AKSZ}. It has been shown that the bosonic, classical  part of the membrane action is given as follows:
\be \label{eq:csmaction}
S[X,A,F]=\int_{\S_3}  \! \! F_i\w\dd X^i+\sfrac 12 \eta_{IJ}A^I\w\dd A^J-\rho^i{}_{I}(X)A^I\w F_i+\sfrac 16T_{IJK}(X)A^I\w A^J\w A^K,
\ee
where  $i=1,\dots,d$ is the target space index and $I=1,\dots,2d$ the pullback bundle index. Here we have maps $X=(X^i):\S_3\to M$, 1-forms $A\in \Omega^1(\S_3,X^{\ast}E)$, and an auxiliary 2-form $F\in \Omega^2(\S_3,X^{\ast}T^{\ast}M)$.
The symmetric bilinear form corresponds to the $O(d,d)$ invariant metric:
\be \label{eta}
\eta=(\eta_{IJ})=\begin{pmatrix}
	0 & 1_d \\ 
	1_d & 0
\nn \end{pmatrix} ,
\ee
while functions $\rho(X)$ and $T(X)$ are related to the anchor and twist of the Courant algebroid, the latter generating a generalised Wess-Zumino term. The full definition of the Courant algebroid can be found  in \cite{LWX}, while  structures relevant for our analysis  shall be defined in Sect. 5. As discussed in Refs.\cite{Ikeda, Roytenberg:2006qz, proc}  the gauge transformations of the CSM mediated by two gauge parameters define a first-stage reducible gauge symmetry, and the  algebra of transformations closes only on-shell.  In the following we shall describe this rich gauge structure using the  $L_\infty$-algebra framework. 

\subsection{Maurer-Cartan homotopy action}

In order to construct the $L_\infty$-algebra for the Courant sigma model we have to define relevant physical fields and assign an appropriate $L_\infty$ grading to each one. 
Starting from the action \eqref{eq:csmaction}, we choose  $\{X^i, A^I, F_i\}$ as physical fields. In order to associate higher products to each of the terms in the action we interpret  functions  $\rho$ and  $T$ as infinite perturbative expansions in field $X$ via their Taylor series. Therefore the action has infinitely many interaction terms:\footnote{In order to make such long expressions more manageable the shorthand $f(0)\equiv f$ and $\partial f\big|_0\equiv\partial f$ for any function $f$ of $X$ evaluated at 0 is used, additionally, the exterior product of forms will be implied with $\w$ suppressed. If the full function is meant the argument will be explicitly written.}
 \begin{align}\label{eq:csmactiontaylor}
 S[X,A,F]=\int_{\S_3}  \! \! &F_i\dd X^i+\sfrac 12 \eta_{IJ}A^I\dd A^J-\rho^i{}_{I}A^I F_i-X^{i_1}\partial_{i_1}\rho^i{}_{I}A^I F_i-{}\nonumber\\
  \phantom{\int_{\S_3}  \! \! }&-\sfrac 1 2 X^{i_1}X^{i_2}\partial_{i_1}\partial_{i_2}\rho^i{}_{I}A^I F_i-\cdots-\sfrac{1}{n!}X^{i_1}\cdots X^{i_n}\partial_{i_1}\cdots\partial_{i_n}\rho^i{}_{I}A^I F_i-\cdots+{}\nonumber\\
 \phantom{\int_{\S_3}  \! \! }&+\sfrac 16T_{IJK}A^I A^J A^K+\sfrac 16X^{i_1}\partial_{i_1}T_{IJK}A^I A^J A^K+{}\nonumber\\
 \phantom{\int_{\S_3}  \! \! }&+\sfrac 1 {12}X^{i_1}X^{i_2}\partial_{i_1}\partial_{i_2}T_{IJK}A^I A^J A^K+{}\nonumber\\
 \phantom{\int_{\S_3}  \! \! }&+\cdots+\sfrac 1 {6\cdot n!} X^{i_1}\cdots X^{i_n}\partial_{i_1}\cdots\partial_{i_n}T_{IJK}A^I A^J A^K+\cdots,
 \end{align}
  all of which must be integrated into the $L_\infty$ picture. Because of this expansion there will not be a finite number of higher products as every interaction term (of which there are infinitely many) will require a unique product. Recalling the construction in \cite{OB} we know all physical fields $a=X+A+F$ are elements of $\mathsf{L'}_1$, their \emph{curvatures} or equations of motion $f=f_X+f_A+f_F$ of $\mathsf{L'}_2$, gauge parameters $c_0=\epsilon+t$ of $\mathsf{L'}_0$ and level 1 parameter $c_{-1}=v$ of $\mathsf{L'}_{-1}$.\footnote{For completeness, here we anticipate appearance of ``ghost for ghost'' gauge parameter $v$ that becomes relevant only in BRST setting.} Thus the complex on which we base the construction of the $L_\infty$    is therefore:
 \be\label{eq:lprimecomplex}
\cdots\to \mathsf{L'}_{-1}\xrightarrow{\mu'_1}\mathsf{L'}_{0}\xrightarrow{\mu'_1}\mathsf{L'}_{1}\xrightarrow{\mu'_1}\mathsf{L'}_{2}\to\cdots.
 \ee
 To obtain expressions in the form of action \eqref{eq:csmaction} or \eqref{eq:csmactiontaylor} we must decompose $\mathsf{L'}$ into the de Rham part $\Omega^\bullet(\Sigma_3)$ and the algebraic $L_\infty$ part $\mathsf{L}$. Since we have three types of fields we shall define $\mathsf{L}$ with three homogeneous subspaces to form the following complex:
 \begin{align}\label{eq:chaincsm}
\mathsf{L}_{-1}\xrightarrow{\mu_1}\mathsf{L}_{0}\xrightarrow{\mu_1}\mathsf{L}_{1}.
\end{align}
Therefore, the classical field content is given below:
\be
\begin{split}
a=X+A+F&\in\Omega^0(M,\mathsf{L}_1)\oplus\Omega^1(M,\mathsf{L}_0)\oplus\Omega^2(M,\mathsf{L}_{-1}),\\
c_0=\epsilon+ t&\in \Omega^0(M,\mathsf{L}_0)\oplus\Omega^1(M,\mathsf{L}_{-1}),\\
c_{-1}=v&\in\Omega^0(M,\mathsf{L}_{-1}),\label{eq:fields}
\end{split}
\ee
and shown in the following table:
\[
\begin{array}{ccccccccccccc}
\cdots & \overset{\mu'_1}{\rightarrow} & \mathsf{L'}_{-1}&\overset{\mu'_1}{\rightarrow}  & \mathsf{L'}_0&\overset{\mu'_1}{\rightarrow}  & \mathsf{L'}_1 &\overset{\mu'_1}{\to} & \mathsf{L'}_2 &\overset{\mu'_1}{\to} & \mathsf{L'}_3&\overset{\mu'_1}{\rightarrow}  & \cdots \\
& &{\scriptstyle{\rm h.\;gauge}}& &{\scriptstyle{\rm gauge}}&  &{\scriptstyle{\rm physical}}&  &{\scriptstyle{\rm equations}} &  &{\scriptstyle{\rm Noether}} &&\\[-2mm]
& &{\scriptstyle{\rm parameters}}& &{\scriptstyle{\rm parameters}}&  &{\scriptstyle{ \rm fields}}&  &{\scriptstyle{\rm of\;motion}} &  &{\scriptstyle{\rm identities}}&&\\[-2mm]
  & &  & & &  & & & & & &&\\
 & \mathsf{L}_{-1} &v_i & & t_i & &  F_i &  & \mathcal{D}F_i & &  &&\\
 &\hspace{-5mm}{\scriptstyle\mu_1} \downarrow   &  & & & & & & &  \\
& \mathsf{L}_0& &  & \epsilon^I &  & A^I & & \mathcal{D}A^I & &&& \\
&\hspace{-5mm}{\scriptstyle\mu_1}\downarrow & & & & & & & & &&\\
& \mathsf{L}_1 &  & & &  & X^i &  &  \mathcal{D}X^i & & &&
\end{array}\]

Once we placed fields in their appropriate homogeneous subspaces, we have to define all the products. Note that the physical field $A$  lives in the pullback bundle $X^*E$  and there is no natural structure defined on its sections; in particular the bracket of sections  $A$ is not the Courant bracket. Thus one can think of $L_\infty$-products as defining relations for the relevant structures on  sections of the pullback bundle. Our selection for the nonvanishing higher products of $\mathsf{L}=\mathsf{L}_1\oplus\mathsf{L}_0\oplus\mathsf{L}_{-1}$ is:
\begin{align}
\mathsf{L}_1&\ni & \mu_n(l_{(1)1},\ldots,l_{(1)n-1},l_{(0)})&=l^{i_1}_{(1)1}\cdots l^{i_{n-1}}_{(1)n-1}\partial_{i_1}\cdots\partial_{i_{n-1}}\rho^i{}_Il^I_{(0)} ,\label{eq:mu10}\\
\mathsf{L}_0&\ni & \mu_n(l_{(1)1},\ldots,l_{(1)n-1},l_{(-1)})&=-l^{i_1}_{(1)1}\cdots l^{i_{n-1}}_{(1)n-1}\partial_{i_1}\cdots\partial_{i_{n-1}}\rho^i{}_Jl_{(-1)i}\eta^{IJ} ,\nn\\
\mathsf{L}_{-1}&\ni & \mu_m(l_{(1)1},\ldots,l_{(1)m-2},l_{(-1)},l_{(0)})&=-l^{i_1}_{(1)1}\cdots l^{i_{m-2}}_{(1)m-2}\partial_{i_1}\cdots\partial_{i_{m-2}}\partial_i\rho^j{}_Il_{(-1)j}l^I_{(0)} ,\nn\\
\mathsf{L}_0&\ni & \mu_m(l_{(1)1},\ldots,l_{(1)m-2},l_{(0)1},l_{(0)2})&=l^{i_1}_{(1)1}\cdots l^{i_{m-2}}_{(1)m-2}\partial_{i_1}\cdots\partial_{i_{m-2}}T_{JKL}l_{(0)1}^Kl_{(0)2}^L \eta^{IJ} ,\nn\\
\mathsf{L}_{-1}&\ni & \mu_r(l_{(1)1},\ldots,l_{(1)r-3},l_{(0)1},l_{(0)2},l_{(0)3})&=l^{i_1}_{(1)1}\cdots l^{i_{r-3}}_{(1)r-3}\partial_{i_1}\cdots\partial_{i_{r-3}}\partial_iT_{IJK}l_{(0)1}^I l_{(0)2}^Jl_{(0)3}^K ,\nn
\end{align}
where $n\geqslant1$, $m\geqslant2$ and $r\geqslant3$, and $l_{(i)}\in\mathsf{L}_i$.\footnote{We defined the products by comparing the general expression for the MC action \eqref{eq:mcaction} with \eqref{eq:csmactiontaylor}, but they can also be obtained from the (expanded) homological vector $Q$ ~\cite{Voronov} defined for a Courant algebroid described as  a QP2-manifold in \cite{dee}.} Tensoring products \eqref{eq:mu10} with the de Rham complex produces for the physical fields:
 \begin{align}
 \mu'_1(X)&=\dd X,\nn\\
 \mu'_1(A)&=\dd A-\mu_1(A),\label{eq:mu1x}\\
 \mu'_1(F)&=\dd F+\mu_1(F),\nn\\
 \intertext{and in general with $n\geqslant 2$ and $m\geqslant 3$:}
 \mu'_n(X_1,\ldots,X_{n-1},A)&=-X^{i_1}_1\cdots X^{i_{n-1}}_{n-1}\partial_{i_1}\cdots\partial_{i_{n-1}}\rho^i{}_IA^I,\nn\\
 \mu'_n(X_1,\ldots,X_{n-1},F)&=-X^{i_1}_1\cdots X^{i_{n-1}}_{n-1}\partial_{i_1}\cdots\partial_{i_{n-1}}\rho^i{}_JF_i\eta^{IJ},\nn\\
 \mu'_n(X_1,\ldots,X_{n-2},A_1,A_2)&=X^{i_1}_1\cdots X^{i_{n-2}}_{n-2}\partial_{i_1}\cdots\partial_{i_{n-2}}T_{JKL}A_1^K A_2^L \eta^{IJ},\label{eq:munxaaa}\\
 \mu'_n(X_1,\ldots,X_{n-2},F,A)&=X^{i_1}_1\cdots X^{i_{n-2}}_{n-2}\partial_{i_1}\cdots\partial_{i_{n-2}}\partial_i\rho^j{}_IA^IF_j,\nn\\
 \mu'_m(X_1,\ldots,X_{m-3},A_1,A_2,A_3)&=-X^{i_1}_1\cdots X^{i_{m-3}}_{m-3}\partial_{i_1}\cdots\partial_{i_{m-3}}\partial_iT_{IJK}A_1^I A_2^J A_3^K.\nn
 \end{align}
 
 Finally, as we are given the action \eqref{eq:csmactiontaylor}, we need to define a consistent inner product. We find that the following choice 
 \begin{align}\label{eq:inner}
 \langle l_{(0)1}, l_{(0)2}\rangle&\equiv\eta_{IJ}l_{(0)1}^I l_{(0)2}^J, & \langle l_{(1)}, l_{(-1)}\rangle&\equiv l_{(1)}^i l_{(-1)i}, & \langle l_{(-1)}, l_{(1)}\rangle&\equiv -l_{(-1)}^i l_{(1)i}.
 \end{align}
satisfies the  cyclicity condition \eqref{eq:cyclic}. Moreover, it defines the pairing on the pullback bundle $X^*E$. Having defined all ingredients we proceed to  calculate the Maurer-Cartan action using \eqref{eq:mcaction}. From the combinatorics of the decomposition of $\mu'_n(a=X+A+F,\ldots,a=X+A+F)$ and the fact that all higher products of physical fields are symmetric in $\mu'$ we obtain:
 \begin{align}
 \mu'_n(a,\ldots,a)&=n\mu'_n(X,\ldots,X,A)+n\mu'_n(X,\ldots,X,F)+\sfrac 1 2 n(n-1)\mu'_n(X,\ldots,X,A,A)+{}\nonumber\\
 &\phantom{\,=\,}+n(n-1)\mu'_n(x,\ldots,X,F,A)+\sfrac{1}{3!}n(n-1)(n-2)\mu'_n(X,\ldots,X,A,A,A).\label{eq:combinatorics}
 \end{align}
 Making use of this decomposition and \eqref{eq:muprimei} the Maurer-Cartan homotopy action \eqref{eq:mcaction}:
 \begin{align*}
 S_{\mathrm{MC}}[X,A,F]&=\langle \dd X,F\rangle+\sfrac 1 2 \langle A,\dd A\rangle+\sum_{n\geqslant 0}\sfrac{1}{n!}\langle A,\mu_{n+1}(X,\ldots,X,F)\rangle+{}\\
 &\phantom{\,=\,} +\sfrac 1 6 \sum_{n\geqslant0}\sfrac{1}{n!}\langle A,\mu_{n+2}(X,\ldots,X,A,A)\rangle,
 \end{align*}
 defined by products \eqref{eq:mu10} and the cyclic inner product \eqref{eq:inner}, indeed corresponds to the desired action \eqref{eq:csmaction} or \eqref{eq:csmactiontaylor}. One can also calculate the equations of motion \eqref{eq:mceom}:
 \begin{align*}
 f_{1}&=\dd X-\sum_{n\geqslant0}\sfrac{1}{n!}\mu_{n+1}(X,\ldots,X,A),\\
 f_0&=\dd A+\sum_{n\geqslant0}\sfrac{1}{n!}\mu_{n+1}(X,\ldots,X,F)+\sfrac 1 2\sum_{n\geqslant0}\sfrac{1}{n!}\mu_{n+2}(X,\ldots,X,A,A),\\
 f_{-1}&=\dd F-\sum_{n\geqslant0}\sfrac{1}{n!}\mu_{n+2}(X,\ldots,X,F,A)-\sfrac{1}{3!}\sum_{n\geqslant0}\sfrac{1}{n!}\mu_{n+3}(X,\ldots,X,A,A,A).
 \end{align*}
 With the aid of \eqref{eq:mu10}  it becomes obvious that these indeed correspond to the equations of motion of action \eqref{eq:csmaction} or \eqref{eq:csmactiontaylor}:
 \begin{align}
 \mathcal{D}X^i&=\dd X^i-\rho^i{}_J(X)A^J,\label{eq:Feom}\\
 \mathcal{D}A^I&=\dd A^I-\eta^{IJ}\rho^j{}_J(X)F_j+\sfrac 1 2 \eta^{IJ}T_{JKL}(X)A^K A^L,\\
 \mathcal{D}F_i&=\dd F_i+\partial_i\rho^j{}_J(X)A^JF_j-\sfrac{1}{3!}\partial_iT_{IJK}(X)A^IA^J A^K\label{eq:Xeom}.
 \end{align}
 
 \subsection{Gauge symmetry}
 Moving on now to the spaces $\mathsf{L'}_0$ and $\mathsf{L'}_{-1}$, they contain gauge parameters $\epsilon$ and $t$, and $v$, respectively. The gauge variations \eqref{eq:gaugea} are:
 \begin{align*}
 \delta_{(\epsilon,t)}X&=\sum_{n\geqslant 0}\sfrac{1}{n!}\mu_{n+1}(X,\ldots,X,\epsilon),\\
 \delta_{(\epsilon,t)}A&=\dd\epsilon-\sum_{n\geqslant 0}\sfrac{1}{n!}\mu_{n+1}(X,\ldots,X,t)+\sum_{n\geqslant 0}\sfrac{1}{n!}\mu_{n+2}(X,\ldots,X,A,\epsilon),\\
  \delta_{(\epsilon,t)}F&=\dd t+\sum_{n\geqslant 0}\sfrac{1}{n!}\mu_{n+2}(X,\ldots,X,A,t)+\sum_{n\geqslant 0}\sfrac{1}{n!}\mu_{n+2}(X,\ldots,X,F,\epsilon)+{}\\
  &\phantom{\,=\,}+ \sfrac 1 2 \sum_{n\geqslant 0}\sfrac{1}{n!}\mu_{n+3}(X,\ldots,X,A,A,\epsilon),
 \end{align*}
 which corresponds to the standard gauge variations of the Courant sigma model \cite{Ikeda}:
 \begin{align}
 \delta_{(\epsilon,t)}X^i&=\rho^i{}_J(X)\epsilon^J,\label{eq:xvar}\\
 \delta_{(\epsilon,t)}A^I&=\dd\epsilon^I+\eta^{IJ}\rho^j{}_J(X)t_j+\eta^{IJ}T_{JKL}(X)A^K\epsilon^L,\\
 \delta_{(\epsilon,t)}F_i&=\dd t_i+\partial_i\rho^j{}_J(X)A^Jt_j-\partial_i\rho^j{}_J(X)\epsilon
 ^JF_j+\sfrac 1 2 \partial_iT_{IJK}(X)A^IA^J\epsilon^K.\label{eq:Fvar}
 \end{align}
 We are left with the higher level 1 gauge transformations of parameters $\epsilon$ and $t$:
 \begin{align*}
 \delta_{v}\epsilon&=\sum_{n\geqslant 0}\sfrac{1}{n!}\mu_{n+1}(X,\ldots,X,v),\\
 \delta_{v}t&=\dd v+\sum_{n\geqslant 0}\sfrac{1}{n!}\mu_{n+2}(X,\ldots,X,A,v),
 \end{align*}
 which, using \eqref{eq:mu10}, give:
 \begin{align*}
 \delta_{v}\epsilon^I&=-\eta^{IJ}\rho^j{}_J(X)v_j,\\
 \delta_{v}t_i&=\dd v_i+\partial_i\rho^j{}_J(X)v_jA^J.
 \end{align*}
 It is perhaps worth noting  that while these higher ``transformations'' have no meaning in classical field theory and are just algebraic, they suggest more structure which becomes relevant as we move towards quantisation and the BV-BRST formulation of the model.
 
\subsection{Homotopy identities}\label{subsec:csmhomotopy}
One crucial element of our construction of $\mathsf{L}$ has been so far omitted, namely the homotopy Jacobi identities \eqref{eq:homotopyjac} of products \eqref{eq:mu10}. To see what constraints these conditions place on our theory we shall calculate them explicitly now. As a reminder we state the first three identities of \eqref{eq:homotopyjac} again:
\begin{align*}
\mu_1(\mu_1(l))&=0\\
\mu_1(\mu_2(l_1,l_2))&=\mu_2(\mu_1(l_1),l_2)-(-1)^{|l_1||l_2|}\mu_2(\mu_1(l_2),l_1)\\
\mu_1(\mu_3(l_1,l_2,l_3))&=\mu_2(\mu_2(l_1,l_2),l_3)-(-1)^{|l_2||l_3|}\mu_2(\mu_2(l_1,l_3),l_2)+{}\\
&\phantom{\,=\,}+(-1)^{|l_1|(|l_2|+|l_3|)}\mu_2(\mu_2(l_2,l_3),l_1)-\mu_3(\mu_1(l_1),l_2,l_3)+{}\\
&\phantom{\,=\,}+(-1)^{|l_1||l_2|}\mu_3(\mu_1(l_2),l_1,l_3)-(-1)^{|l_3|(|l_1|+|l_2|)}\mu_3(\mu_1(l_3),l_1,l_2).
\end{align*}
The products \eqref{eq:mu10}  give the following conditions for some of the first three identities:\footnote{Underlined indices are not antisymmetrised.}
  \begin{align*}
 i&=1: & l&=l_{(1)}: & & & \text{trivial}&\\
 & & l&=l_{(0)}: & & & \text{trivial}&\\
 & & l&=l_{(-1)}: & &\Rightarrow &  \eta^{IJ}\rho^i{}_I\rho^j{}_J=0&\\
 i&=2: & l_{1,2}&=l_{(1)1},l_{(1)2}: & & & \text{trivial}&\\
 & & l_{1,2}&=l_{(1)},l_{(0)}: & &  & \text{trivial}&\\
 & & l_{1,2}&=l_{(1)},l_{(-1)}: & &\Rightarrow  & \partial_{i}(\eta^{IJ}\rho^j{}_I\rho^k{}_J)=0&\\
 & & l_{1,2}&=l_{(0)1},l_{(0)2}: & &\Rightarrow & 2\rho^j{}_{[I}\partial_{\underline{j}}\rho^i{}_{J]}-\rho^i{}_M\eta^{ML}T_{LIJ}=0&\\
 & & l_{1,2}&=l_{(0)},l_{(-1)}: & &\Rightarrow & 2\rho^j{}_{[I}\partial_{\underline{j}}\rho^i{}_{J]}-\rho^i{}_M\eta^{ML}T_{LIJ}=0&\\
 & & l_{1,2}&=l_{(-1)1},l_{(-1)2}: & &\Rightarrow & \partial_{i}(\eta^{IJ}\rho^j{}_I\rho^k{}_J)=0&\\
 i&=3: & l_{1,2,3}&=l_{(0)1},l_{(0)2},l_{(0)3}: & &\Rightarrow & 3\rho^i{}_{[A}\partial_{\underline{i}}T_{BC]J}-\rho^i{}_J\partial_{i}T_{ABC}-3T_{JK[A}\eta^{KM}T_{BC]M}=0&
 \end{align*}
 In general, for arbitrary $n\geqslant1$ we have at most seven nontrivial homotopy identities of which only three are unique, as we show explicitly in Appendix \ref{app:csmhomotopy}. 
 These three sets of homotopy conditions for the higher products are actually all terms in the Taylor expansions of the axioms of the Courant algebroid (taking $l_{(1)i}=X_i$) of which each order must hold separately. Classically, these follow from the gauge invariance of the action \eqref{eq:csmaction}:
 \bea\label{flux}
&&  \eta^{IJ}\rho^i{}_I(X)\rho^j{}_J(X)=0,\nn\\
 && 2\rho^j{}_{[I}(X)\partial_{\underline{j}}\rho^i{}_{J]}(X)-\rho^i{}_M(X)\eta^{ML}T_{LIJ}(X)=0,\\
 && 3\rho^i{}_{[A}(X)\partial_{\underline{i}}T_{BC]J}(X)-\rho^i{}_J(X)\partial_{i}T_{ABC}(X)-3T_{JK[A}(X)\eta^{KM}T_{BC]M}(X)=0.\nn
 \eea
As expected,  all these identities `live' in the $\mathsf{L'}_3$ space, the space of Noether identities of classical field theory, c.f.\cite{fulp,brano}. 
 In the worldsheet approach to non-geometric string backgrounds, these conditions were seen to originate from generalised Wess-Zumino terms giving expressions for fluxes and their Bianchi identities~\cite{Hal, Mylonas, ChJL, Watamura, p12}.  The `non-geometric' here means that the string background fields  defined over overlapping open neighbourhoods are patched using diffeomorphisms, and gauge and  T-duality transformations. In the ${L_\infty}$-algebra  formulation of the CSM, the expressions for the corresponding fluxes and their Bianchi identities result from higher gauge symmetries encoded in the homotopy relations. 
  The benefit of this interpretation is that one knows how to  extend the obtained  classical expressions  to the full BV-BRST action, as we discuss in the next section. 
 
 \section{$L_\infty$ for BV-BRST  Courant sigma model}
 
 Moving away from our classical results and going towards quantisation one encounters the need for BRST symmetry. BRST at the most trivial level is the promotion of gauge parameters to (propagating) \emph{ghost} fields. For certain gauge theories this is not enough and one must introduce more fields (often called antifields) to be able to quantise, this is the Batalin-Vilkovisky procedure. The Courant sigma model is one such theory as its gauge algebra is open (see e.g.~\cite{proc}), one could also see this from the existence of a higher gauge parameter in the classical $L_\infty$ picture of the previous sections. In the next two sections we shall give an overview of how to discover BV-BRST within $L_{\infty}$ following \cite{brano} and then use this to calculate the generalised BRST transformations and BV action for the Courant sigma model.
 \subsection{BV and $L_\infty$-algebras}
 To introduce the Batalin-Vilkovisky formalism into a homotopy theory we consider a cyclic $L_{\infty}$-algebra $(\mathsf{L},\mu_i,\langle  \,\cdot\,, \,\cdot\,  \rangle_{\mathsf{L}})$ with $|\langle  \,\cdot\,, \,\cdot\,  \rangle_{\mathsf{L}}|=-3$. The truncation (in that the complex stops at 1) of the $L_{\infty}$-algebra to include fields and ghosts:
	\be\label{eq:brstlprime}
	\mathsf{L}_{\mathrm{BRST}}=\bigoplus_{i\leqslant1}\mathsf{L}_i,
	\ee
is the {BRST complex}. In terms of our classical picture of the previous section this is the field content of complex \eqref{eq:lprimecomplex} truncated after $\mathsf{L'}_1$ as higher spaces do not correspond to fields if one reinterprets (higher) gauge parameters as ghost fields. As is customary we will write the fields as contracted coordinate functions,\footnote{Given a graded vector space $\mathsf{V}$, coordinate functions are maps:
	\begin{align*}
	\xi^\alpha:\mathsf{V}\to\mathbb{R}\quad\text{such that}\quad\xi^\alpha(v)=\xi^\alpha(v^\beta\tau_\beta)=v^\alpha,
	\end{align*}
	with degrees: $|\xi^\alpha|=-|v|\equiv-|\tau_\alpha|$.
	We use the convention that the degree-shifted graded vector space $\mathsf{V}[k]$ has vectors of degree $|v|-k$, which implies the coordinate functions will then be of degree $|\xi^\alpha|=-|\tau_\alpha|+k$.\label{foot:contracteddeg}
}
\[
\xi=\xi^\alpha\tau_\alpha,\qquad \xi^\alpha\in\mathcal{C}^{\infty}(\mathsf{L}[1]),\;\tau_\alpha\in\mathsf{L}.
\]
They are useful as one can write $L_\infty$-algebra objects in a basis independent form. Therefore, the $L_\infty$-algebra of interest is:
\[
\hat{\mathsf{L}}_{\mathrm{BRST}}\equiv\mathcal{C}^\infty(\mathsf{L}_{\mathrm{BRST}}[1])\otimes\mathsf{L}_{\mathrm{BRST}}.
\]

A few slight modifications must be made for this tensor product algebra; $\mathcal{C}^\infty(\mathsf{L}[1])$ can be understood as a differential graded commutative algebra with trivial differential therefore $\hat{\mu}_1$ only has the second term in \eqref{eq:muprime1}, and since the pairing on $\mathsf{L}$ itself can have a non-zero degree $k=|\langle\,\cdot\,, \,\cdot\, \rangle_{\mathsf{L}}|$, $\zeta_1,\zeta_2\in\mathcal{C}^\infty(\mathsf{L}[1])$ will additionally graded commute with the pairing producing a second sign. The cyclic inner product thus decomposes as:
\be
\langle \zeta_1\otimes l_1,\zeta_2\otimes l_2\rangle_{\mathsf{\hat L}}=(-1)^{k(|\zeta_1|+|\zeta_2|)+|\zeta_2||l_1|}(\zeta_1\zeta_2)\langle l_1,l_2\rangle_{\mathsf{L}}.\label{eq:hatpairing}
\ee
The $\hat{\mathsf{L}}$ degree\footnote{$\hat{\mathsf{L}}$ as a tensor product space has elements with a bi-degree, the \emph{ghost number} or degree in $\mathcal{C}^\infty(\mathsf{L}_{\mathrm{BRST}}[1])$ and $L_\infty$-degree or degree in $\mathsf{L}$, and as an $L_\infty$-algebra they have a single $\hat{\mathsf{L}}$ degree being the sum of its bi-degrees.} of all contracted coordinate functions of fields is then 1 (see footnote \ref{foot:contracteddeg}). Therefore we can combine all gauge fields and ghosts into a single contracted coordinate function \emph{superfield}:
\[
\mathsf{a}_{\mathrm{BRST}}\equiv a+\sum_{i\geqslant 0}c_{-i}.
\]
The action of the BRST operator $Q_{\mathrm{BRST}}$ is then:
\be
Q_{\mathrm{BRST}}\mathsf{a}_{\mathrm{BRST}}=-\sum_{i\geqslant1}\frac{1}{i!}\hat{\mu}_i(\mathsf{a}_{\mathrm{BRST}},\ldots,\mathsf{a}_{\mathrm{BRST}}).\label{eq:brstoperator}
\ee
In this truncated $L_\infty$-algebra however, $Q_{\mathrm{BRST}}$ is in general only nilpotent up to the Maurer-Cartan equation. Since this violation of nilpotency stems precisely from the truncation, by extending this complex to all homogeneous subspaces (essentially letting $i\in\mathbb{Z}$ in \eqref{eq:brstlprime}) one regains $Q$ as a homological operator again, this is the BV complex. 

In the language of field theory this is simply a reformulation of the requirement for BV in open algebra gauge cases, in other words this is nothing more than the addition of antifields for every physical and ghost field. We associate these {antifields}\footnote{A dagger in superscript will always denote the corresponding antifield.} $a^\dagger$ and $c_{-k}^\dagger$ $(k\geqslant0)$ to each gauge or ghost field. The BV \emph{superfield} is then also by extension a combination of all the gauge and ghost fields, and antifields:
\[
\mathsf{a}\equiv a+a^\dagger+\sum_{i\geqslant 0}(c_{-i}+c_{-i}^\dagger).
\]
The curvature of $\mathsf{a}$ is as in any $L_\infty$-algebra given by:
 \begin{align}
 \mathsf{f}=\sum_{i\geqslant1}\frac{1}{i!}\hat{\mu}_i(\mathsf{a},\ldots,\mathsf{a}).
 \end{align}
 Realising the operator $Q_\text{BV}$ simply reduces to $Q_\text{BRST}$ when $\mathsf{a}$ is truncated, implies $Q_\text{BV}$ is defined as in \eqref{eq:brstoperator} with $\mathsf{a}$ the full BV superfield. This means we have the action of $Q_\text{BV}$ given by 
 \begin{align}\label{eq:bvtransf}
 Q_{\mathrm{BV}}\mathsf{a}=-\mathsf{f},\qquad\qquad Q_{\mathrm{BV}}\mathsf{f}=0.
 \end{align}
 From this it is obvious that $Q_{\mathrm{BV}}\mathsf{a}$ will contain the classical gauge variations:
 \[Q_{\mathrm{BV}}a=\delta_{c_0}a+\cdots\qquad Q_{\mathrm{BV}}c_{-k}=(-1)^{k+1}\delta_{c_{-k-1}}c_{-k}+\cdots\]
 where $k\geqslant 0$ and equations of motion:
 \[Q_{\mathrm{BV}}a^\dagger=-f+\cdots.\]
	The function $S_{\mathrm{BV}}$ on $\mathcal{F}_{\mathrm{BV}}$ defined as:
	\begin{align}
	S_{\mathrm{BV}}[\mathsf{a}]\equiv\sum_{i\geqslant0}\frac{1}{(i+1)!}\langle \mathsf{a},\hat{\mu}_i(\mathsf{a},\ldots,\mathsf{a})\rangle_{\hat{\mathsf{L}}},\label{eq:bvaction}
	\end{align}
	is the BV extension of \eqref{eq:mcaction} called the \textbf{Maurer-Cartan-Batalin-Vilkovisky action} or BV action for short.
For details see \cite{brano}.

\subsection{Maurer-Cartan BV for the CSM}
 In BRST quantisation gauge parameters $\epsilon$ and $t$ become ghost fields, however, since the gauge algebra is a reducible one the higher gauge field becomes a scalar ghost-for-ghost field: $v$.\footnote{Denoted the same as the gauge parameter they originate from.} As stated in the previous section to complete the BV formalism we extend this BRST complex with the antifields corresponding to each BRST field. Following the construction of \cite{brano} we assign these fields to the three ${L_\infty}$-algebra $(\mathsf{L},\mu_i)$ spaces as shown in table \ref{table:degrees}. 
 
 {\centering
\tikzset{node style ge/.style={circle}}

\begin{tikzpicture}[baseline=(A.center)]
\tikzset{BarreStyle/.style =   {opacity=.3,line width=6 mm,line cap=round,color=#1}}
\tikzset{SigneAbove/.style =   {above right,,opacity=0.9, text=#1}}
\tikzset{SigneAboveRight/.style =   {above right,,opacity=0.9,text=#1}}
\matrix (A) [matrix of math nodes, nodes = {node style ge},,column sep=9 mm, row sep=9 mm] 
{\Omega^\bullet & \mathsf{L}_{-1} & \mathsf{L}_{0} & \mathsf{L}_{1}\\
	0 & v & \epsilon & X  \\
	1 & t & A & F^\dagger  \\
	2 & F & A^\dagger & t^\dagger \\
	3 & X^\dagger & \epsilon^\dagger & v^\dagger \\
};

\draw [BarreStyle=lightgray]  (A-2-2.south west) to (A-2-2.north east);
\draw [BarreStyle=darkgray]  (A-3-2.south west) to (A-2-3.north east);
\draw [BarreStyle=lightgray]  (A-4-2.south west) to (A-2-4.north east);
\draw [BarreStyle=darkgray]  (A-5-2.south west) to (A-3-4.north east);
\draw [BarreStyle=lightgray]  (A-5-3.south west) to (A-4-4.north east);
\draw [BarreStyle=darkgray]  (A-5-4.south west) to (A-5-4.north east);
\draw[opacity=0] (-2.9,-0.15) -- (-0.85,1.75) node[sloped, above=5pt, right=17pt, pos=1, text=darkgray, opacity=0.9]{\scriptsize$\gh c_{-1}=2$};
\draw[opacity=0] (-2.9,-0.15) -- (-0.85,1.75) node[sloped, below=4pt, right=17pt, pos=1, text=darkgray, opacity=0.9]{\scriptsize$|c_{-1}|_{\mathsf{L'}}=-1$};
\draw[opacity=0] (-2.9,-2) -- (1.2,1.75) node[sloped, above=5pt, right=15pt, pos=1, text=darkgray, opacity=0.9]{\scriptsize$\gh c_0=1$};
\draw[opacity=0] (-2.9,-2) -- (1.2,1.75) node[sloped, below=4pt, right=15pt, pos=1, text=darkgray, opacity=0.9]{\scriptsize$|c_{0}|_{\mathsf{L'}}=0$};
\draw[opacity=0] (-2.9,-3.85) -- (3.03,1.75) node[sloped, above=5pt, right=20pt, pos=1, text=darkgray, opacity=0.9]{\scriptsize$\gh a=0$};
\draw[opacity=0] (-2.9,-3.85) -- (3.03,1.75) node[sloped, below=4pt, right=20pt, pos=1, text=darkgray, opacity=0.9]{\scriptsize$|a|_{\mathsf{L'}}=1$};
\draw[opacity=0] (-0.85,-3.85) -- (3.03,-0.05) node[sloped, above=5pt, right=21pt, pos=1, text=darkgray, opacity=0.9]{\scriptsize$\gh a^\dagger=-1$};
\draw[opacity=0] (-0.85,-3.85) -- (3.03,-0.05) node[sloped, below=4pt, right=21pt, pos=1, text=darkgray, opacity=0.9]{\scriptsize$|a^\dagger|_{\mathsf{L'}}=2$};
\draw[opacity=0] (1.05,-3.9) -- (3.0,-1.95) node[sloped, above=5pt, right=20pt, pos=1, text=darkgray, opacity=0.9]{\scriptsize$\gh c_0^\dagger=-2$};
\draw[opacity=0] (1.05,-3.9) -- (3.0,-1.95) node[sloped, below=4pt, right=20pt, pos=1, text=darkgray, opacity=0.9]{\scriptsize$|c_0^\dagger|_{\mathsf{L'}}=3$};
\draw[opacity=0] (2.7,-4.2) -- (3.3,-3.58) node[sloped, above=5pt, right=10pt, pos=1, text=darkgray, opacity=0.9]{\scriptsize$\gh c_{-1}=-3$};
\draw[opacity=0] (2.7,-4.2) -- (3.3,-3.58) node[sloped, below=4pt, right=10pt, pos=1, text=darkgray, opacity=0.9]{\scriptsize$|c_{-1}^\dagger|_{\mathsf{L'}}=4$};
\end{tikzpicture}
\captionof{table}{Degrees of fields.}\label{table:degrees}
}

 Therefore the complete BV field content is:\footnote{The number in the square bracket indicates the fields' ghost degree.}
\be
\begin{split}
a=X+A+F&\in\Omega^0(M,\mathsf{L}_1)[0]\oplus\Omega^1(M,\mathsf{L}_0)[0]\oplus\Omega^2(M,\mathsf{L}_{-1})[0],\\
a^\dagger =X^\dagger+A^\dagger+F^\dagger&\in\Omega^3(M,\mathsf{L}_{-1})[-1]\oplus\Omega^2(M,\mathsf{L}_0)[-1]\oplus\Omega^1(M,\mathsf{L}_{1})[-1],\\
c_0=\epsilon+ t&\in \Omega^0(M,\mathsf{L}_0)[1]\oplus\Omega^1(M,\mathsf{L}_{-1})[1],\\
c_{-1}=v&\in\Omega^0(M,\mathsf{L}_{-1})[2],\\
c_0^\dagger=\epsilon^\dagger+t^\dagger&\in\Omega^3(M,\mathsf{L}_0)[-2]\oplus\Omega^2(M,\mathsf{L}_1)[-2],\\
c_{-1}^\dagger=v^\dagger&\in\Omega^3(M,\mathsf{L}_1)[-3],\label{eq:BVfields}
\end{split}
\ee
where fields with the same ghost number are collected. Additionally, these $\mathsf{L'}$ fields can also be combined into the BV \emph{superfield} $\mathsf{a}$:\footnote{There is a slight abuse of notation here as $a$, $a^\dagger$, $c_0$, $c_0^\dagger$, $c_{-1}$ and $c_{-1}^\dagger$ have two meanings: they are elements of $\mathsf{L'}$ as was the case in \eqref{eq:BVfields}, however, in $\mathsf{a}$ they imply being elements of $\mathcal{C}^\infty(\mathsf{L'}[1])\otimes\mathsf{L'}$ as well, this is for reasons of brevity and the desired meaning should be clear from context. See also footnote \ref{foot:ghostbases}\label{foot:abuseofnotation}}
\[
\mathsf{a}=a+a^\dagger+c_0+c_0^\dagger+c_{-1}+c_{-1}^\dagger,
\]
of $\hat{\mathsf{L}}$-degree 1. Using the properties of cyclic inner product for $L_\infty$-algebras \eqref{eq:cyclic} and tensored $L_\infty$-algebras  \eqref{eq:hatpairing} and the combinatorics of decomposing $\mathsf{a}$ just as in \eqref{eq:combinatorics}, the BV action \eqref{eq:bvaction} becomes:\footnote{As stated in footnote \ref{foot:abuseofnotation} the explicit writing of ghost bases will be suppressed. Therefore, in all expressions in which it is not explicitly written it will be assumed $\mu'_i(l'_1,\ldots,l'_i)$ stands for $\zeta_1\cdots\zeta_i\;\mu'_i(l'_1,\ldots,l'_i)$.\label{foot:ghostbases}}
\begin{align*}
S_{BV}&=\int_{\Sigma_3}\!\!\langle\dd X,F\rangle+\sfrac 1 2 \langle A,\dd A\rangle+{}\\
&\phantom{\,=\,\int_{\Sigma_3}\!\!\langle\dd X}+\sum_{n=0}^\infty\frac{1}{n!}\Big(\langle 
F,\mu_{n+1}(X,\ldots,X,A)\rangle+ \sfrac{1}{6  }\langle 
A,\mu_{n+2}(X,\ldots,X,A,A)\rangle\Big)+{}\\
&\phantom{\,=\,}+\int_{\Sigma_3}-\langle F^\dagger
,\dd t\rangle-\langle A^\dagger, \dd\epsilon\rangle-\langle t^\dagger,\dd v\rangle+\sum_{n=0}^{\infty}  \frac{1}{n!}\Big(- \langle 
v^\dagger,\mu_{n+2}(X,\ldots,X,v,\epsilon)\rangle-{}\\
&\hphantom{\,=\,+\int_{\Sigma_3}}-  \langle t^\dagger,\mu_{n+3}(X,\ldots,X,F^\dagger,v,\epsilon)\rangle- \sfrac{1}{2  }\langle A,\mu_{n+3}(X,\ldots,X,F^\dagger,F^\dagger,v)\rangle+{}\\
&\hphantom{\,=\,+\int_{\Sigma_3}}+ \sfrac{1}{6  }\langle F^\dagger,\mu_{n+4}(X,\ldots,X,F^\dagger,F^\dagger,v,\epsilon)\rangle+
\langle t^\dagger,\mu_{n+2}(X,\ldots,X,v,A)\rangle+{}\\
&\hphantom{\,=\,+\int_{\Sigma_3}}+  \langle A^\dagger,\mu_{n+2}(X,\ldots,X,F^\dagger,v)\rangle+  \langle 
\epsilon^\dagger,\mu_{n+1}(X,\ldots,X,v)\rangle+{}\\
&\hphantom{\,=\,+\int_{\Sigma_3}}+ \sfrac{1}{2  }\langle 
t,\mu_{n+3}(X,\ldots,X,F^\dagger,F^\dagger,\epsilon)\rangle-  \langle t,\mu_{n+2}(X,\ldots,X,F^\dagger,A)\rangle+{}\\
&\hphantom{\,=\,+\int_{\Sigma_3}}+  \langle A^\dagger,\mu_{n+1}(X,\ldots,X,t)\rangle-  \langle 
t^\dagger,\mu_{n+2}(X,\ldots,X,t,\epsilon)\rangle-{}\\
&\hphantom{\,=\,+\int_{\Sigma_3}}-  \langle
F,\mu_{n+2}(X,\ldots,X,F^\dagger,\epsilon)\rangle-  \langle 
X^\dagger,\mu_{n+1}(X,\ldots,X,\epsilon)\rangle-{}\\
&\hphantom{\,=\,+\int_{\Sigma_3}}- \sfrac{1}{6  }\langle v^\dagger,\mu_{n+3}(X,\ldots,X,\epsilon,\epsilon,\epsilon)\rangle- \sfrac{1}{6  }\langle 
t^\dagger,\mu_{n+4}(X,\ldots,X,F^\dagger,\epsilon,\epsilon,\epsilon)
\rangle+{}\\
&\hphantom{\,=\,+\int_{\Sigma_3}}+ \sfrac{1}{2  }\langle t^\dagger,\mu_{n+3}(X,\ldots,X,A,\epsilon,\epsilon)\rangle+ \sfrac{1}{2  }\langle 
\epsilon^\dagger,\mu_{n+2}(X,\ldots,X,\epsilon,\epsilon)\rangle-{}\\
&\hphantom{\,=\,+\int_{\Sigma_3}}- \sfrac{1}{4  }\langle A,\mu_{n+4}(X,\ldots,X,F^\dagger,F^\dagger,\epsilon,\epsilon)\rangle+ \sfrac{1}{2  }\langle 
A^\dagger,\mu_{n+3}(X,\ldots,X,F^\dagger,\epsilon,\epsilon)\rangle+{}\\
&\hphantom{\,=\,+\int_{\Sigma_3}}+ \sfrac{1}{2  }\langle 
A,\mu_{n+3}(X,\ldots,X,F^\dagger,A,\epsilon)\rangle-  \langle 
A^\dagger,\mu_{n+2}(X,\ldots,X,A,\epsilon)\rangle+{}\\
&\hphantom{\,=\,+\int_{\Sigma_3}}+ \sfrac{1}{36  }\langle F^\dagger,\mu_{n+5}(X,\ldots,X,F^\dagger,F^\dagger,\epsilon,\epsilon,\epsilon)\rangle\Big).
\end{align*}
An equivalent procedure by use of \eqref{eq:brstoperator} produces the following components of curvature $\mathsf{f}$ or BV-BRST transformations \eqref{eq:bvtransf}:
\begin{align*}
Q_{BV}v&=\sum_{n=0}^\infty  \frac{1}{n!}\Big(  \mu_{n+2}(X,\ldots,X,v,\epsilon)+ \sfrac{1}{6  }\mu_{n+3}(X,\ldots,X,\epsilon,\epsilon,\epsilon)\Big),\\
Q_{BV}\epsilon&=\sum_{n=0}^\infty  \frac{1}{n!} \Big(- \mu_{n+1}(X,\ldots,X,v)-  \sfrac{1}{2  }\mu_{n+2}(X,\ldots,X,\epsilon,\epsilon)\Big),\\
Q_{BV}X&=\sum_{n=0}^\infty  \frac{1}{n!}  \mu_{n+1}(X,\ldots,X,\epsilon),\\
Q_{BV}t&=-\dd v+\sum_{n=0}^\infty  \frac{1}{n!}\Big(  \mu_{n+2}(X,\ldots,X,v,A)-   \mu_{n+2}(X,\ldots,X,t,\epsilon)-{} \\
&\phantom{\,=\,}- \mu_{n+3}(X,\ldots,X,F^\dagger,v,\epsilon)+  \sfrac{1}{2  }\mu_{n+3}(X,\ldots,X,A,\epsilon,\epsilon)-{}\\
&\phantom{\,=\,}-  \sfrac{1}{6  }\mu_{n+4}(X,\ldots,X,F^\dagger,\epsilon,\epsilon,\epsilon)\Big),\\
Q_{BV}A&=\dd\epsilon +\sum_{n=0}^\infty  \frac{1}{n!}\Big( - \mu_{n+1}(X,\ldots,X,t)+   \mu_{n+2}(X,\ldots,X,A,\epsilon)-{}\\
&\phantom{\,=\,}-   \mu_{n+2}(X,\ldots,X,F^\dagger,v)-  \sfrac{1}{2  }\mu_{n+3}(X,\ldots,X,F^\dagger,\epsilon,\epsilon)\Big),\\
Q_{BV}F^\dagger&=-\dd X+\sum_{n=0}^\infty  \frac{1}{n!} \Big( \mu_{n+1}(X,\ldots,X,A)-   \mu_{n+2}(X,\ldots,X,F^\dagger,\epsilon)\Big),\\
Q_{BV}F&=\dd t+\sum_{n=0}^\infty  \frac{1}{n!} \Big(  \mu_{n+2}(X,\ldots,X,t,A)+   \mu_{n+2}(X,\ldots,X,F,\epsilon)+{}\\
&\phantom{\,=\,\dd t+\sum_{n=0} \mu_{n+2}}+\sfrac{1}{2  }\mu_{n+3}(X,\ldots,X,A,A,\epsilon)+{}\\
&\phantom{\,=\,}+   \mu_{n+3}(X,\ldots,X,F^\dagger,v,A)-   \mu_{n+3}(X,\ldots,X,F^\dagger,t,\epsilon)+{}\\
&\phantom{\,=\,}+   \mu_{n+3}(X,\ldots,X,t^\dagger,v,\epsilon)+  \sfrac{1}{2  }\mu_{n+3}(X,\ldots,X,A^\dagger,\epsilon,\epsilon)+\mu_{n+2}(X,\ldots,X,v,A^\dagger)+{}\\
&\phantom{\,=\,}+  \sfrac{1}{2  }\mu_{n+4}(X,\ldots,X,F^\dagger,A,\epsilon,\epsilon)-  \sfrac{1}{12  }\mu_{n+5}(X,\ldots,X,F^\dagger,F^\dagger,\epsilon,\epsilon,\epsilon)+{}\\
&\phantom{\,=\,}+  \sfrac{1}{6  }\mu_{n+4}(X,\ldots,X,t^\dagger,\epsilon,\epsilon,\epsilon) -  \sfrac{1}{2  }\mu_{n+4}(X,\ldots,X,F^\dagger,F^\dagger,v,\epsilon)\Big),\\
Q_{BV}A^\dagger&=-\dd A+\sum_{n=0}^\infty  \frac{1}{n!} \Big( -  \sfrac{1}{2  }\mu_{n+2}(X,\ldots,X,A,A)- \mu_{n+1}(X,\ldots,X,F)-{}\\
&\phantom{\,=\,}-   \mu_{n+2}(X,\ldots,X,t^\dagger,v)-   \mu_{n+2}(X,\ldots,X,A^\dagger,\epsilon)+   \mu_{n+2}(X,\ldots,X,F^\dagger,t)-{}\\
&\phantom{\,=\,}-   \mu_{n+3}(X,\ldots,X,F^\dagger,A,\epsilon)+  \sfrac{1}{4  }\mu_{n+4}(X,\ldots,X,F^\dagger,F^\dagger,\epsilon,\epsilon)+{}\\
&\phantom{\,=\,}+  \sfrac{1}{2  }\mu_{n+3}(X,\ldots,X,F^\dagger,F^\dagger,v)-  \sfrac{1}{2  }\mu_{n+3}(X,\ldots,X,t^\dagger,\epsilon,\epsilon)\Big),\\
Q_{BV}t^\dagger&=-\dd F^\dagger+\sum_{n=0}^\infty  \frac{1}{n!} \Big(  \mu_{n+1}(X,\ldots,X,A^\dagger)+   \mu_{n+2}(X,\ldots,X,F^\dagger,A)-{}\\
&\phantom{\,=\,}-  \sfrac{1}{2  }\mu_{n+3}(X,\ldots,X,F^\dagger,F^\dagger,\epsilon)+   \mu_{n+2}(X,\ldots,X,t^\dagger,\epsilon)\Big),\\
Q_{BV}X^\dagger&=-\dd F+\sum_{n=0}^\infty  \frac{1}{n!} \Big(  \mu_{n+2}(X,\ldots,X,F,A)+ \sfrac{1}{6  }\mu_{n+3}(X,\ldots,X,A,A,A)  -{} \\
&\phantom{\,=\,}- \mu_{n+2}(X,\ldots,X,X^\dagger,\epsilon)-   \mu_{n+3}(X,\ldots,X,F^\dagger,v,A^\dagger)+  \sfrac{1}{2  }\mu_{n+3}(X,\ldots,X,\epsilon^\dagger,\epsilon,\epsilon)-{}\\
&\phantom{\,=\,}-   \mu_{n+3}(X,\ldots,X,F^\dagger,F,\epsilon)-  \sfrac{1}{2  }\mu_{n+4}(X,\ldots,X,F^\dagger,F^\dagger,v,A)+{}\\
&\phantom{\,=\,}+  \sfrac{1}{6  }\mu_{n+5}(X,\ldots,X,F^\dagger,F^\dagger,F^\dagger,v,\epsilon)+   \mu_{n+3}(X,\ldots,X,t^\dagger,v,A)-{}\\
&\phantom{\,=\,}-   \mu_{n+4}(X,\ldots,X,F^\dagger,t^\dagger,v,\epsilon)-   \mu_{n+3}(X,\ldots,X,v^\dagger,v,\epsilon)+\mu_{n+2}(X,\ldots,X,v,\epsilon^\dagger)-{}\\
&\phantom{\,=\,}-   \mu_{n+3}(X,\ldots,X,A^\dagger,A,\epsilon)-   \mu_{n+3}(X,\ldots,X,F^\dagger,t,A)-{}\\
&\phantom{\,=\,}-  \sfrac{1}{2  }\mu_{n+4}(X,\ldots,X,F^\dagger,A^\dagger,\epsilon,\epsilon)-  \sfrac{1}{4  }\mu_{n+5}(X,\ldots,X,F^\dagger,F^\dagger,A,\epsilon,\epsilon)+{}\\
&\phantom{\,=\,}+  \sfrac{1}{36  }\mu_{n+6}(X,\ldots,X,F^\dagger,F^\dagger,F^\dagger,\epsilon,\epsilon,\epsilon)+  \sfrac{1}{2  }\mu_{n+4}(X,\ldots,X,t^\dagger,A,\epsilon,\epsilon)-{}\\
&\phantom{\,=\,}-  \sfrac{1}{6  }\mu_{n+5}(X,\ldots,X,F^\dagger,t^\dagger,\epsilon,\epsilon,\epsilon)-  \sfrac{1}{6  }\mu_{n+4}(X,\ldots,X,v^\dagger,\epsilon,\epsilon,\epsilon)-{}\\
&\phantom{\,=\,}-   \mu_{n+2}(X,\ldots,X,t,A^\dagger)+  \sfrac{1}{2  }\mu_{n+4}(X,\ldots,X,F^\dagger,F^\dagger,t,\epsilon)-{}\\
&\phantom{\,=\,}-  \sfrac{1}{2  }\mu_{n+4}(X,\ldots,X,F^\dagger,A,A,\epsilon)-   \mu_{n+3}(X,\ldots,X,t^\dagger,t,\epsilon)\Big),\\
Q_{BV}\epsilon^\dagger&=\dd A^\dagger+\sum_{n=0}^\infty  \frac{1}{n!} \Big(  -\mu_{n+1}(X,\ldots,X,X^\dagger)-   \mu_{n+2}(X,\ldots,X,F^\dagger,F)+{}\\
&\phantom{\,=\,}+  \sfrac{1}{6  }\mu_{n+4}(X,\ldots,X,F^\dagger,F^\dagger,F^\dagger,v)-   \mu_{n+2}(X,\ldots,X,t^\dagger,t)-{}\\
&\phantom{\,=\,}-   \mu_{n+2}(X,\ldots,X,v^\dagger,v)+   \mu_{n+2}(X,\ldots,X,\epsilon^\dagger,\epsilon)-{}\\
&\phantom{\,=\,}-  \sfrac{1}{2  }\mu_{n+3}(X,\ldots,X,F^\dagger,A,A)-   \mu_{n+3}(X,\ldots,X,F^\dagger,A^\dagger,\epsilon)-{}\\
&\phantom{\,=\,}-   \mu_{n+2}(X,\ldots,X,A^\dagger,A)+   \mu_{n+3}(X,\ldots,X,t^\dagger,A,\epsilon)-{}\\
&\phantom{\,=\,}-  \sfrac{1}{2  }\mu_{n+3}(X,\ldots,X,v^\dagger,\epsilon,\epsilon)+  \sfrac{1}{2  }\mu_{n+3}(X,\ldots,X,F^\dagger,F^\dagger,t)-{}\\
&\phantom{\,=\,}-  \sfrac{1}{2  }\mu_{n+4}(X,\ldots,X,F^\dagger,t^\dagger,\epsilon,\epsilon)-   \mu_{n+3}(X,\ldots,X,F^\dagger,t^\dagger,v)+{}\\
&\phantom{\,=\,}+  \sfrac{1}{12  }\mu_{n+5}(X,\ldots,X,F^\dagger,F^\dagger,F^\dagger,\epsilon,\epsilon)-  \sfrac{1}{2  }\mu_{n+4}(X,\ldots,X,F^\dagger,F^\dagger,A,\epsilon)\Big),\\
Q_{BV}v^\dagger&=-\dd t^\dagger+\sum_{n=0}^\infty  \frac{1}{n!} \Big(  \mu_{n+1}(X,\ldots,X,\epsilon^\dagger)-   \mu_{n+2}(X,\ldots,X,F^\dagger,A^\dagger)+{}\\
&\phantom{\,=\,}+  \sfrac{1}{6  }\mu_{n+4}(X,\ldots,X,F^\dagger,F^\dagger,F^\dagger,\epsilon)+   \mu_{n+2}(X,\ldots,X,t^\dagger,A)-{}\\
&\phantom{\,=\,}-   \mu_{n+2}(X,\ldots,X,v^\dagger,\epsilon)-  \sfrac{1}{2  }\mu_{n+3}(X,\ldots,X,F^\dagger,F^\dagger,A)-{}\\
&\phantom{\,=\,}-   \mu_{n+3}(X,\ldots,X,F^\dagger,t^\dagger,\epsilon)\Big).
\end{align*}
Introducing our selection for the higher products \eqref{eq:mu10}  and inner product \eqref{eq:inner}, and then resumming the Taylor expansions produces the full BV action (as obtained by the AKSZ procedure in \cite{Roytenberg:2006qz}). The explicit results are given in Appendix \ref{app:BVappendix}. 

So far, we have constructed the cyclic $L_\infty$-algebra underlying the CSM, and obtained, by tensoring it with the de Rham complex, the dynamics of the model -- the action, equations of motion and gauge transformations. Finally, since the CSM has an open gauge algebra we proceeded to finalise the theory with its BV description obtained in the framework of $L_\infty$-algebras by introducing a third algebra which includes the ghost degrees. Using this framework enabled us to obtain the exact BV-BRST transformations of both fields and antifields in our theory (physical and ghost). In the next section we would like to relate this $L_\infty$-algebra corresponding to the Courant sigma model with the 2-term  $L_\infty$-algebra for a Courant algebroid in Ref.\cite{Roytenberg:1998vn}.

 \section{$L_\infty$-morphisms}

Roytenberg has shown \cite{Roytenberg:2006qz} that given the data of a Courant algebroid one can uniquely construct the corresponding  Courant sigma model. Moreover, Roytenberg and Weinstein  showed  \cite{Roytenberg:1998vn} that a Courant algebroid  can be described as 2-term $L_\infty$-algebra. This naturally raises the question of the relation between the $L_\infty$-algebra we constructed for the CSM and the one defined in Ref.\cite{Roytenberg:1998vn}. 
We first recall the definition of $L_\infty$-morphisms. A morphism between two $L_\infty$-algebras $(\tilde{\mathsf{L}},\tilde{\mu}_i)$ and $(\mathsf{L},\mu_i)$ is a collection of homogeneous maps $\phi_i:\tilde{\mathsf{L}}\times\hdots\times \tilde{\mathsf{L}}\to \mathsf{L}$ of degree $1-i$ for $ i \in \mathbb{N}$ which are multilinear and totally graded anti-symmetric and obey:
\begin{align}\label{morph}
&\sum_{j+k=i}\sum_{\sigma\in \mathrm{Sh}(j;i)}(-1)^{k}\chi(\sigma;l_1,\ldots,l_i)\phi_{k+1}(\tilde{\mu}_j(l_{\s(1)},\ldots,l_{\s(j)}),l_{\s(j+1)},\ldots,l_{\s(i)})=\nn\\
&= \sum_{j=1}^i\sfrac {1}{j!}  \sum_{k_1+\cdots+k_j=i}\sum_{\sigma\in \mathrm{Sh}(k_1,\ldots,k_{j-1};i)}\chi(\sigma;l_1,\ldots,l_i)\zeta(\sigma;l_1,\ldots,l_i)\times \\
&\phantom{\,=\,}{}\times \mu_j(\phi_{k_1}(l_{\s(1)},\ldots,l_{\s(k_1)}),\ldots,\phi_{k_j}(l_{\s(k_1+\hdots+k_{j-1}+1)},\ldots, l_{\s(i)})) ,
\nn
\end{align}
where $\chi(\sigma;l_1,\ldots,l_i)$  is the graded Koszul sign and $\zeta(\sigma;l_1,\ldots,l_i)$ for a $(k_1,\ldots,k_{j.1};i)$-shuffle $\s$ is given by
\bea
\zeta(\sigma;l_1,\ldots,l_i)=(-1)^{\sum_{1\leqslant m<n\leqslant j}k_mk_n+\sum_{m=1}^{j-1}k_m(j-m)+\sum_{m=2}^j(1-k_m)
	\sum_{k=1}^{k_1+\hdots+k_{m-1}}|l_{\s(k)}|} .\nn\eea
	
\subsection{$L_\infty$-algebra for a Courant algebroid}\label{subsec:Linftyextension}

We start from a Courant algebroid  as an $L_\infty$-algebra of \cite{Roytenberg:1998vn} concentrated in  two spaces:
\[\widetilde{\mathsf{L}}_{-1}\overset{\mathcal{D}}{\to}\widetilde{\mathsf{L}}_0,\]
with $\widetilde{\mathsf{L}}_{-1}=C^\infty(M)$ and $\widetilde{\mathsf{L}}_0=\Gamma(E)$, and non-vanishing products:\footnote{We dropped the space of constants $\widetilde{\mathsf{L}}_{-2}$  as they play no role in the following and signs are adjusted to match our conventions as explained in Appendix \ref{app:conventions}.}
\begin{align}
\label{CA}
\tilde{\mu}_1(f)&={\cal D}f ,\nn\\
\tilde{\mu}_2(e_1,e_2)&=[e_1,e_2]_C ,\nn\\
\tilde{\mu}_2(e,f)&=\langle e,{\cal D}f\rangle ,\nn\\
\tilde{\mu}_3(e_1,e_2,e_3)&={\cal N}(e_1,e_2,e_3) ,
\end{align}
with $f\in C^\infty(M)$ and $e\in \Gamma(E)$. The maps are defined in terms of structures on a Courant algebroid, i.e.: map ${\cal D}:  C^\infty(M)\to  \Gamma(E)$, a skew-symmetric bracket on sections of bundle $E$ over manifold $M$, a symmetric  bilinear form $\langle\;,\;\rangle$  and  tensor ${\cal N}$ representing the obstruction to the Jacobi identity of the bracket, 
${\cal N}(e_1,e_2,e_3)=\sfrac 13\langle [e_1,e_2]_C,e_3\rangle+{\rm cycl.}$. Explicit expressions will be given in next subsections when we construct the morphisms.  In the definition of a Courant algebroid (see e.g.\cite{Roytenberg:1998vn}) these structures satisfy certain compatibility conditions which are in the $L_\infty$ formulation given by homotopy relations. 

To make the connection with our sigma model algebra \eqref{eq:chaincsm} and \eqref{eq:mu10}, we must extend\footnote{In nomenclature of Ref.\cite{OB} we have to extend pure gauge algebra to algebra including additional field which will correspond to the field $X$ in CSM. The reason is that the CSM algebra is field dependent.}   the chain complex of this algebra by an additional space of degree 1, $\widetilde{\mathsf{L}}_1=T_pM$:
\begin{equation}\label{eq:extendedcomplex}
\widetilde{\mathsf{L}}_{-1}\overset{\mathcal{D}}{\to}\widetilde{\mathsf{L}}_0\overset{\tilde{\rho}}{\to}\widetilde{\mathsf{L}}_1,
\end{equation}
where,
\begin{equation}\label{eq:mu1e}
\tilde{\mu}_1(e)=\tilde{\rho}(e)\big|_p,
\end{equation}
is the map $\tilde\rho:E\to TM$ and $p\in M$ is a point on manifold $M$. We will denote elements of $\widetilde{\mathsf{L}}_1$ by $h\in T_pM$. Calculation of the homotopy identities \eqref{eq:homotopyjac} (for details see Appendix \ref{sec:CAhomotopy}) provides the minimal extension to the higher products \eqref{CA} necessary to make \eqref{eq:extendedcomplex} an $L_\infty$-algebra:
\begin{equation}
\tilde{\mu}_n(h_1,\ldots,h_{n-1},e)=h_1^{i_1}\cdots h_{n-1}^{i_{n-1}}\tilde\partial_{i_{1}}\cdots\tilde\partial_{i_{i-1}}(\tilde{\rho}(e)^i)\big|_p,\qquad\qquad n\in\mathbb{N},
\end{equation}
where the basis of $T_pM$ is the one induced by coordinates $x^i$ of a coordinate patch $U\subset M$ that contains point $p$ such that $x^i(p)=0$. It is important to note that this extended algebra also corresponds to the Courant algebroid as did the original $L_2$ formulation since no new properties, other than the Courant algebroid axioms, were needed.

\subsection{$L_\infty$-morphism from CA to CSM}

Now, we shall construct the morphism $\phi$ to our CSM algebra in a pointwise fashion i.e. $\phi:\tilde{\mathsf{L}}\times\cdots\times\tilde{\mathsf{L}}\to \mathsf{L}\big|_{X(\sigma)=p}$ for $\sigma\in\Sigma_3$, since the Courant sigma model is only locally defined. We begin analogously to the extension procedure of the previous paragraph given explicitly in Appendix \ref{sec:CAhomotopy}. The construction will follow orders of $i$ in \eqref{morph}.\\\\
\underline{$i=1$}\\
To begin, we start by the lowest order of \eqref{morph} which encompasses two non-trivial conditions:
\begin{align*}
\phi_1(\tilde{\mu}_1(e))&=\mu_1(\phi_1(e)),\\
\phi_1(\tilde{\mu}_1(f))&=\mu_1(\phi_1(f)).
\end{align*}
By use of the first equation in \eqref{eq:mu10} and \eqref{eq:mu1e} the first relation gives:
\[\phi_1(\tilde{\rho}(e)^i\big|_p)=\rho^i{}_I\phi_1(e)^I,\]
whereas the second  equation in \eqref{eq:mu10} and \eqref{CA} produce:
\[\phi_1(\mathcal{D}f)^I=-\eta^{IJ}\rho^i{}_J\phi_1(f)_i.\]
These two relations imply $\phi_1$ to be:
\begin{align}
\phi_1(h)&=X^*h,\\
\phi_1(e)&=X^*e\big|_p,\\
\phi_1(f)&=-X^*\tilde\dd f\big|_p,
\end{align}
where we used $\langle{\cal D}f,e\rangle=\sfrac 12 \rho(e)f$ and  $\rho^i{}_J\equiv \rho^i{}_J(\{X^j\}=0)=\rho^i{}_J(\{X^*x^j\}=0)$ and $x^i(p)=0$.\\\\
\underline{$i=2$}\\
In this case we have four non-trivial morphism conditions. The first is $(l_1,l_2)=(e_1,e_2)$:
\begin{align*}
-\phi_2(\tilde{\mu}_1(e_1),e_2)+\phi_2(\tilde{\mu}(e_2),e_1)+\phi_1(\tilde{\mu}_2(e_1,e_2))&=\mu_1(\phi_2(e_1,e_2))+\mu_2(\phi_1(e_1),\phi_1(e_2))\\
-\phi_2(\tilde{\rho}(e_1)\big|_p,e_2)^I+\phi_2(\tilde\rho(e_2)\big|_p,e_1)^I+X^*([e_1,e_2]_C)^I\big|_p&=-\eta^{IJ}\rho^i{}_J\phi_2(e_1,e_2)_i+{}\\
&\phantom{\,=\,}+\eta^{IJ}T_{JKL}(X^*e_1\big|_p)^K(X^*e_2\big|_p)^L.
\end{align*}
By comparison with the Courant bracket:
\begin{align}\label{eq:courantbracket}
([e_1,e_2]_C)^A&=\tilde\r^a{}_B(e_1^B\tilde\partial_ae_2^A-e_2^B\tilde\partial_ae_1^A)-{}\\
&\phantom{\,=\,}-\sfrac 12\tilde\r^a{}_C(e_1^B\tilde\partial_ae_{2B}-e_2^B\tilde\partial_ae_{1B})\tilde\eta^{AC}+\tilde\eta^{AB}\tilde T_{BCD}e_1^Ce_2^D ,\nn
\end{align}
we fix two $\phi_2$ maps:
\begin{align}
\phi_2(h,e)^I&=X^*(h^i\tilde{\partial}_ie^I)\big|_p,\\
\phi_2(e_1,e_2)&=X^*(\eta_{IJ}e^I_{[1}\tilde{\dd}e^J_{2]})\big|_p.
\end{align}
The second condition corresponding to $(l_1,l_2)=(e,f)$ is:
\begin{align*}
\phi_1(\tilde{\mu}_2(e,f))-\phi_2(\tilde{\mu}_1(e),f)+\phi_2(\tilde{\mu}_1(f),e)&=\mu_2(\phi_1(e),\phi_1(f))\\
-X^*(\tilde\dd\langle e,\mathcal{D}f\rangle)_i\big|_p-\phi_2(\tilde{\rho}(e)\big|_p,f)_i+\sfrac 1 2 (X^*(\mathcal{D}_I f\tilde{\dd}e^I)\big|_p)_i&-\sfrac 1 2 (X^*( e^I\tilde{\dd}\mathcal{D}_I f)\big|_p)_i=\\
&=-\partial_i\rho^j{}_I(X^*\tilde{\dd}f)_j\big|_p(X^*e)^I\big|_p,
\end{align*}
from which we read off:
\begin{equation}
\phi_2(h,f)_i=-X^*(h^j\tilde{\partial}_j\tilde{\partial}_i f)\big|_p.
\end{equation}
For the third combination of elements we take $(l_1,l_2)=(e,h)$:
\begin{align*}
\phi_1(\tilde{\mu}_2(e,h))-\phi_2(\tilde{\mu}_1(e),h)&=\mu_1(\phi_2(e,h))+\mu_2(\phi_1(e),\phi_1(h))\\
X^*(h^j\tilde{\partial}_j(\tilde{\rho}(e)^i)\big|_p)+\phi_2(\tilde{\rho}(e)\big|_p,h)^i&=\rho^i{}_I(X^*(h^j\tilde{\partial}_j e)\big|_p)^I+(X^*h)^j\partial_j\rho^i{}_I(X^*e\big|_p)^I,
\end{align*}
that allows us to set:
\begin{equation}
\phi_2(h_1,h_2)=0.
\end{equation}
The final relation with $(l_1,l_2)=(f,h)$:
\[-\phi_2(\tilde{\mu}_1(f),h)=\mu_1(\phi_2(f,h))+\mu_2(\phi_1(f),\phi_1(h)),\]
is just a consistency check.\\\\
\underline{$i=3$}\\
Out of the five non-trivial conditions that exist for $i\geqslant3$ we begin with the combination $(l_1,l_2,l_3)=(e_1,e_2,e_3)$:
\begin{align*}
\sfrac 1 3 \phi_1(\tilde{\mu}_3(e_1,e_2,e_3))&-\phi_2(\tilde{\mu}_2(e_1,e_2),e_3)+\phi_3(\tilde{\mu}_1(e_1),e_2,e_3)+\text{cyclic}={}\\
&=\sfrac 1 3 \mu_3(\phi_1(e_1),\phi_1(e_2),\phi_1(e_3))-\mu_2(\phi_1(e_3),\phi_2(e_1,e_2))+\text{cyclic}.
\end{align*}
This condition implies:
\begin{equation}
\phi_3(h,e_1,e_2)_i=X^*(h^j\tilde{\partial}_j(\eta^{IJ}e^I_{[1}\tilde{\partial}_ie^J_{2]}))\big|_p.
\end{equation}
The next combination of elements $(l_1,l_2,l_3)=(e_1,e_2,h)$ gives:
\begin{align*}
\phi_3&(\tilde{\mu}_1(e_1),e_2,h)+\phi_2(\tilde{\mu}_2(e_1,h),e_2)+\sfrac 1 2 \phi_2(\tilde{\mu}_2(e_2,e_1),h)-e_1\leftrightarrow e_2={}\\
&=\sfrac 1 2 \mu_1(\phi_3(e_1,e_2,h))+\mu_2(\phi_1(e_2),\phi_2(e_1,h))+\sfrac 1 2 \mu_2(\phi_1(h),\phi_2(e_1,e_2))+{}\\
&\phantom{\,=\,}+\sfrac 1 2 \mu_3(\phi_1(e_1),\phi_1(e_2),\phi_1(h))-e_1\leftrightarrow e_2.
\end{align*}
Expanding this expression and plugging in the previously set definitions for $\phi$ one can consistently set:
\begin{equation}
\phi_3(e,h_1,h_2)^I=X^*(h_1^{i_1}h_2^{i_2}\tilde{\partial}_{i_{1}}\tilde{\partial}_{i_{2}}e^I)\big|_p.
\end{equation}
The third possibility is for $(l_1,l_2,l_3)=(h,f,e)$:
\begin{align*}
\phi_3(\tilde{\mu}_1(f),h,e)&+\phi_3(\tilde{\mu}_1(e),h,f)+\phi_2(\tilde{\mu}_2(h,e),f)+\phi_2(\tilde{\mu}_2(f,e),h)={}\\
&=\mu_2(\phi_1(f),\phi_2(h,e))-\mu_2(\phi_1(e),\phi_2(h,f))+\mu_3(\phi_1(h),\phi_1(f),\phi_1(e)).
\end{align*}
Analogously to the previous cases, here we can set:
\begin{equation}
\phi_3(h_1,h_2,f)_i=-X^*(h_1^{i_1}h_2^{i_2}\tilde{\partial}_{i_{1}}\tilde{\partial}_{i_{2}}\tilde{\partial}_if)\big|_p.
\end{equation}
There are two more combinations of elements with non-trivial conditions: $(l_1,l_2,l_3)=(h_1,h_2,e)$ and $(l_1,l_2,l_3)=(f,h_1,h_2)$, however, these are simply consistency checks that all other $\phi_3$ can be set to vanish. We state them for completeness:
\begin{align*}
\sfrac 1 2 \phi_3&(\tilde{\mu}_1(e),h_1,h_2)+\phi_2(\tilde{\mu}_2(h_1,e),h_2)+\sfrac 1 2 \phi_1(\tilde{\mu}_3(h_1,h_2,e))+h_1\leftrightarrow h_2={}\\
&=\sfrac 1 2 \mu_1(\phi_3(h_1,h_2,e))+\mu_2(\phi_1(h_1),\phi_2(h_2,e))+\sfrac 1 2 \mu_3(\phi_1(h_1),\phi_1(h_2),\phi_1(e))+h_1\leftrightarrow h_2,\\
\sfrac 1 2 \phi_3&(\tilde{\mu}_1(f),h_1,h_2)+h_1\leftrightarrow h_2={}\\
&=\sfrac 1 2 \mu_1(\phi_3(f,h_1,h_2))+\mu_2(\phi_1(h_1),\phi_2(f,h_2))+\sfrac 1 2 \mu_3(\phi_1(f),\phi_1(h_1),\phi_1(h_2))+h_1\leftrightarrow h_2.
\end{align*}
\underline{$i\geqslant3$}\\
Since the $i=3$ case already gives the most general morphism conditions we make the ansatz for the four possible non-vanishing $\phi_i$ mappings as follows:
\begin{align}
\phi_i(h_1,\ldots,h_{i-1},e)^I&=X^*(h_1^{j_1}\cdots h_{i-1}^{j_{i-1}}\tilde{\partial}_{j_{1}}\cdots\tilde{\partial}_{j_{i-1}}e^I)\big|_p,\\
\phi_i(h_1,\ldots,h_{i-2},e_1,e_2)_j&=X^*(h_1^{j_1}\cdots h_{i-2}^{j_{i-2}}\tilde{\partial}_{j_{1}}\cdots\tilde{\partial}_{j_{i-2}}( \eta_{IJ}e^I_{[1}\tilde{\partial}_j e^J_{2]}))\big|_p,\\
\phi_i(h_1,\ldots,h_{i-1},f)_j&=-X^*(h_1^{j_1}\cdots h_{i-1}^{j_{i-1}}\tilde{\partial}_{j_{1}}\cdots\tilde{\partial}_{j_{i-1}}\tilde{\partial_j}f)\big|_p,\\
\phi_i(h_1,\ldots,h_{i})&=0.
\end{align} First of the five non-trivial conditions corresponds to the choice $(l_1,\ldots,l_i)=(h_1,\ldots,h_{i-1},f)$:
\begin{align*}
(-1)^{i-1}\phi_i(\tilde{\mu}_1(f),h_1,\ldots,h_{i-1})=&\sum_{n=1}^{i}\mu_n(\phi_{i-n+1}(h_1,\ldots,h_{i-n},f),\phi_1(h_{i-n+1}),\ldots,\phi_1(h_{i-1}))+{}\\
&+\text{perm.}
\end{align*}
which is automatically satisfied by use of the ansatz.\footnote{Here ``perm." will indicate all possible unshuffles of $h_1,\ldots,h_{i-1}$.} Next is the combination $(l_1,\ldots,l_i)=(h_1,\ldots,h_{i-1},e)$:
\begin{align*}
\phi_1(\tilde{\mu}_i(h_1,\ldots,h_{i-1},e))=&\sum_{n=1}^{i}\mu_n(\phi_1(h_1),\ldots,\phi_1(h_{n-1}),\phi_{i-n+1}(h_n,\ldots,h_{i-1},e))+{}\\
&+\text{perm.}
\end{align*}
that is also automatically satisfied. The third case is $(l_1,\ldots,l_i)=(h_1,\ldots,h_{i-2},f,e)$:
\begin{align*}
&(-1)^{i-1}\phi_i(\tilde{\mu}_1(f),h_1,\ldots,h_{i-2},e)+\phi_i(\tilde{\mu}_2(f,e),h_1,\ldots,h_{i-2})+{}\\
&+\sum_{n=1}^{i-1}\phi_{i-n+1}(\tilde{\mu}_n(h_1,\ldots,h_{n-1},e),h_n,\ldots,h_{i-2},f)+\text{perm.}={}\\
&=\sum_{m=2}^{i}\sum_{n=1}^{i-m+1}\times{}\\
&\phantom{\,=\,}\times\mu_m(\phi_1(h_1),\ldots,\phi_1(h_{m-2}),\phi_{n}(h_{m-1},\ldots,h_{m+n-3},f),\phi_{i-m-n+2}(h_{m+n-2},\ldots,h_{i-2},e))+{}\\
&\phantom{\,=\,}+\text{perm.}
\end{align*}
which is obviously satisfied after one resums.
The fourth possibility is $(l_1,\ldots,l_i)=(h_1,\ldots,h_{i-2},e_1,e_2)$:
\begin{align*}
(-1)^{i} \phi_{i-1}(\tilde{\mu}_2(e_1,e_2),h_1,\ldots,h_{i-2})&-\Bigg(\sum_{n=1}^{i-1}\phi_{i-n+1}(\tilde{\mu}_n(h_1,\ldots,h_{n-1},e_1),h_n,\ldots,h_{i-2},e_2)+{}\\
&+\text{perm.}-e_1\leftrightarrow e_2\Bigg)={}\\
=\sum_{m=2}^{i}\sum_{n=1}^{i-m+1}\mu_m(\phi_1(h_1),\ldots,\phi_1(h&_{m-2}),\phi_{n}(h_{m-1},\ldots,h_{m+n-3},e_1),\\
&\phi_{i-m-n+2}(h_{m+n-2},\ldots,h_{i-2},e_2)) +\text{perm.}-e_1\leftrightarrow e_2+{}\\
+\sum_{m=1}^{i-1}\mu_m(\phi_1(h_1),\ldots,\phi_1(h_{m-1}&),\phi_{i-m+1}(h_m,\ldots,h_{i-2},e_1,e_2))+\text{perm.}
\end{align*}
This is satisfied by definition of the Courant bracket \eqref{eq:courantbracket}. Finally, the last condition for $(l_1,\ldots,l_i)=(h_1,\ldots,h_{i-3},e_1,e_2,e_3)$ is:
\begin{align*}
&\phi_{i-2}(\tilde{\mu}_3(e_1,e_2,e_3),h_1,\ldots,h_{i-3})+(-1)^{i}\phi_{i-1}(\tilde{\mu}_2(e_1,e_2),h_1,\ldots,h_{i-3},e_3)+\text{cycl.}+{}\\
&\phantom{\,=\,}+\sum_{n=1}^{i-2}\phi_{i-n+1}(\tilde{\mu}_n(h_1,\ldots,h_{n-1},e_1),h_n,\ldots,h_{i-3},e_2,e_3)+\text{perm.}+\text{cycl.}={}\\
&=-\sum_{l=2}^{i-1}\sum_{n=1}^{i-l}\times{}\\
&\phantom{\,=\,-\,}\times\mu_l(\phi_1(h_1),\ldots,\phi_1(h_{l-2}),\phi_{n}(h_{l-1},\ldots,h_{l+n-3},e_1),\phi_{i-l-n+2}(h_{l+n-2},\ldots,h_{i-3},e_2,e_3))-{}\\
&\phantom{\,=\,-\,\times\,}-\text{perm.}-\text{cycl.}+{}\\
&\phantom{\,=\,}+\sum_{l=3}^{i}\sum_{n=1}^{i-l+1}\sum_{m=1}^{i-l-n+2}\mu_l(\phi_1(h_1),\ldots,\phi_1(h_{l-3}),\phi_{n}(h_{l-2},\ldots,h_{l+n-4},e_1),\\
&\phantom{\,=\,-\,\times\,}\phi_{m}(h_{l+n-3},\ldots,h_{m+n+l-5},e_2),\phi_{i-m-n-l+3}(h_{m+n+l-4},\ldots,h_{i-3},e_3))+\text{perm.}+\text{cycl.}
\end{align*}
where ``cycl." indicates all cycles of $e_1,e_2,e_3$. This is satisfied by virtue of cyclicity and the properties of a Courant algebroid as in the $i=3$ case.

As is expected all five conditions are after resumming simply all the Taylor expansion terms as implied by their lowest orders i.e. $l=f$, $l=e$, $(l_1,l_2)=(e,f)$, $(l_1,l_2)=(e_1,e_2)$ and $(l_1,l_2,l_3)=(e_1,e_2,e_3)$.

Thus we have shown how to construct the Courant sigma model starting from the structures of a Courant algebroid encoded in a 2-term $L_\infty$-algebra. One has to extend the pure gauge structure of the 2-term $L_\infty$-algebra of a CA defined in Ref.\cite{Roytenberg:1998vn} to include field dependence. The morphism we constructed produces all brackets defining the CSM $L_\infty$-algebra and thus the Maurer-Cartan equations. However, the MC action requires an additional input and can be constructed only if one can define a consistent bilinear pairing rendering the CSM $L_\infty$-algebra cyclic. 
\paragraph{Acknowledgements.} We thank Athanasios Chatzistavrakidis, Branislav Jur\v{c}o and Christian S\"{a}mann for
helpful discussions. The work  is partially  supported by  the European Union through the European Regional Development Fund -- The Competitiveness and Cohesion Operational Programme (KK.01.1.1.06) and the Croatian Science Foundation under project IP-2019-04-4168.

\appendix 

\section{Conventions}\label{app:conventions}
We provide a short dictionary between conventions of \cite{OB} and \cite{Roytenberg:1998vn}, and \cite{brano} that we use. The first difference is that degrees are inverted as shown in table \ref{table:degconv}
\begin{table}[h]
	\centering
\begin{tabular}{c c c}
$\accentset{\circ}{\mathsf{L}}_{-i}$& $\Leftrightarrow$ &$\mathsf{L}_{i}$\\
$|\accentset{\circ}{\mu_i}|=i-2$& $\Leftrightarrow$ &$|\mu_i|=2-i$
\end{tabular}
\caption{Change of degrees between conventions.}\label{table:degconv}
\end{table}
where by $\circ$ we indicate the conventions of \cite{OB} and \cite{Roytenberg:1998vn}. The second, and much more important, difference is in the homotopy relation that states:
\[
\sum_{j+k=i}\sum_\sigma\chi(\sigma;l_1,\ldots,l_i)(-1)^{kj}\accentset{\circ}{\mu}_{k+1}(\accentset{\circ}{\mu}_j(l_{\sigma(1)},\ldots,l_{\sigma(j)}),l_{\sigma(j+1)},\ldots,l_{\sigma(i)})=0,
\]
notice the $kj$ in the exponent of $-1$ as opposed to just $k$ in \eqref{eq:homotopyjac}. The relation to our convention (other than degree inversion) is given by an additional sign:
\[\mu_j\to(-1)^{\sfrac 1 2 j(j-1)}\accentset{\circ}{\mu}_j,\]
this sign compensates the difference between the homotopy relations (up to an overall sign dependant on $i$ that goes away since the right-hand side is zero). However, as the sign of $\mu$ changes so will the expressions for the Maurer-Cartan equation \eqref{eq:mceom}, homotopy action \eqref{eq:mcaction}:
\begin{align*}
S_{\mathrm{MC}}[a]&=\sum_{i\geqslant1}\frac{(-1)^{\sfrac 1 2 i(i-1)}}{(i+1)!}\langle a,\accentset{\circ}{\mu}'_i(a,\ldots,a)\rangle_{\accentset{\circ}{\mathsf{L}}'},\\
	f&=\sum_{i\geqslant1}\frac{(-1)^{\sfrac 1 2 i(i-1)}}{i!}\accentset{\circ}{\mu}'_i(a,\ldots,a),
\end{align*}
and others, precisely as stated in \cite{OB}.

\section{Homotopy identities of CSM algebra}\label{app:csmhomotopy}
In section \ref{subsec:csmhomotopy} we calculated some of the homotopy identities for $n=1,2,3$. Here we provide the calculation of all homotopy relations for arbitrary $n$. As was stated, there are seven possible combinations of elements that produce nontrivial identities. Each possibility is calculated below.\\
 \begin{itemize}
 \item\underline{$(l_{1},\ldots,l_n)=(l_{(1)1},\ldots,l_{(1)n-1},l_{(-1)})$}
 \begin{align*}
 &\mu_1(\mu_n(l_{(1)1},\ldots,l_{(1)n-1},l_{(-1)}))=\mu_2(\mu_{n-1}(l_{(1)1},\ldots,l_{(1)n-2},l_{(-1)}),l_{(1)n-1}) +\cdots+{}\\
 &+ (-1)^{k+1}\mu_{k+1}(\mu_{n-k}(l_{(1)1},\ldots,l_{(1)n-k-1},l_{(-1)}),l_{(1)n-k},\ldots,l_{(1)n-1})+{}\\
 &+\text{perm.}+\cdots\\
 &-l^{i_1}_{(1)1}\cdots l^{i_{n-1}}_{(1)n-1}\partial_{i_1}\cdots\partial_{i_{n-1}}\rho^j{}_J\eta^{IJ}\rho^i{}_Il_{(-1)j}={}\\
 &=l^{i_{n-1}}_{(1)n-1}\partial_{i_{n-1}}\rho^i{}_I\eta^{IJ}l^{i_{1}}_{(1)1}\cdots l^{i_{n-2}}_{(1)n-2}\partial_{i_1}\cdots\partial_{i_{n-2}}\rho^j{}_Jl_{(-1)j}+\cdots+{}\\
 &+l^{i_{1}}_{(1)1}\cdots l^{i_{n-k-1}}_{(1)n-k-1}\partial_{i_1}\cdots\partial_{i_{n-k-1}}\rho^j{}_Jl^{i_{n-k}}_{(1)n-k}\cdots l^{i_{n-1}}_{(1)n-1}\partial_{i_{n-k}}\cdots\partial_{i_{n-1}}\rho^i{}_I\eta^{IJ}l_{(-1)j}+{}\\
 &+\text{perm.}+\cdots\\
 \end{align*}
 \begin{align}
 &\Downarrow\nonumber\\
 \partial_{i_1}\cdots\partial_{i_{n-1}}(\rho^i{}_I\eta^{IJ}\rho^j{}_J)&=0\label{eq:cond11}
 \end{align}
 Here ``perm.'' denotes all possible permutations of $l_{(1)1},\ldots,l_{(1)n-1}$ that are ordered as required by \eqref{eq:homotopyjac}. All such permutations will have positive sign because the Koszul sign will exactly compensate the permutation sign since all objects are either of degree 1 or $-1$.\\
 \item\underline{$(l_{1},\ldots,l_n)=(l_{(1)1},\ldots,l_{(1)n-2},l_{(-1)1},l_{(-1)2})$}
 \begin{align*}
 &0=\mu_2(\mu_{n-1}(l_{(1)1},\ldots,l_{(1)n-2},l_{(-1)1}),l_{(-1)2}) +\cdots+{}\\
 &+ (-1)^{k+1}\mu_{k+1}(\mu_{n-k}(l_{(1)1},\ldots,l_{(1)n-k-1},l_{(-1)1}),l_{(1)n-k},\ldots,l_{(1)n-2},l_{(-1)2})+{}\\
 &+\text{perm.}+\cdots\\
 &0=-\partial_k\rho^j{}_J\eta^{IJ}l^{i_{1}}_{(1)1}\cdots l^{i_{n-2}}_{(1)n-2}\partial_{i_1}\cdots\partial_{i_{n-2}}\rho^i{}_Il_{(-1)1i}l_{(-1)2j}-\cdots-{}\\
 &-l^{i_{n-k}}_{(1)n-k}\cdots l^{i_{n-2}}_{(1)n-2}\partial_{i_{n-k}}\cdots\partial_{i_{n-2}}\rho^j{}_J\eta^{IJ}l^{i_{1}}_{(1)1}\cdots l^{i_{n-k-1}}_{(1)n-k-1}\partial_{i_{1}}\cdots\partial_{i_{n-k-1}}\rho^i{}_Il_{(-1)1j}l_{(-1)2j}-{}\\
 &-\text{perm.}-\cdots
  \end{align*}
 \begin{align}
 &\Downarrow\nonumber\\
 \partial_{i_1}\cdots\partial_{i_{n-2}}\partial_k(\rho^i{}_I\eta^{IJ}\rho^j{}_J)&=0\label{eq:cond12}
 \end{align}
 In this case ``perm.'' indicates all possible unshuffles of $l_{(1)1},\ldots,l_{(1)n-2}$ and also terms with $l_{(-1)1}$ and $l_{(-1)2}$ swapped. The sign of all terms will be the same for the same reason as above.\\
 \item\underline{$(l_{1},\ldots,l_n)=(l_{(1)1},\ldots,l_{(1)n-2},l_{(0)1},l_{(0)2})$}
 \begin{align*}
 &\mu_1(\mu_n(l_{(1)1},\ldots,l_{(1)n-2},l_{(0)1},l_{(0)2}))=\mu_2(\mu_{n-1}(l_{(1)1},\ldots,l_{(1)n-2},l_{(0)1}),l_{(0)2})-{} \\
 &-l_{(0)1}\leftrightarrow l_{(0)2} +{}\\
 &+\mu_2(\mu_{n-1}(l_{(1)1},\ldots,l_{(1)n-3},l_{(0)1},l_{(0)2}),l_{(1)n-2}) +\cdots+{}\\
 &+ (-1)^{k+1}\mu_{k+1}(\mu_{n-k}(l_{(1)1},\ldots,l_{(1)n-k-2},l_{(0)1},l_{(0)2}),l_{(1)n-k-1},\ldots,l_{(1)n-2})+{}\\
 &+ \mu_{k+1}(\mu_{n-k}(l_{(1)1},\ldots,l_{(1)n-k-1},l_{(0)1}),l_{(1)n-k},\ldots,l_{(1)n-2},l_{(0)2})-l_{(0)1}\leftrightarrow l_{(0)2} +{}\\
 &+\text{perm.}+\cdots\\
 &l^{i_{1}}_{(1)1}\cdots l^{i_{n-2}}_{(1)n-2}\partial_{i_{1}}\cdots\partial_{i_{n-2}}T_{JKL}\eta^{IJ}l^K_{(0)1}l^L_{(0)2}\rho^i{}_I={}\\
 &= l^{i_{1}}_{(1)1}\cdots l^{i_{n-2}}_{(1)n-2}\partial_{i_{1}}\cdots\partial_{i_{n-2}}\rho^j{}_K\partial_j\rho^i{}_Ll^K_{(0)1}l^L_{(0)2}-l_{(0)1}\leftrightarrow l_{(0)2}-{}\\
 &-l^{i_{1}}_{1}\cdots l^{i_{n-3}}_{(1)n-3}\partial_{i_{1}}\cdots\partial_{i_{n-3}}T_{JKL}\eta^{IJ}l^{i_{n-2}}_{(1)n-2}\partial_{i_{n-2}}\rho^i{}_Il^K_{(0)1}l^L_{(0)2}+{}\\
 &+\cdots+{}\\
 &+l^{i_{1}}_{(1)1}\cdots l^{i_{n-k-1}}_{(1)n-k-1}\partial_{i_{1}}\cdots\partial_{i_{n-k-1}}\rho^j{}_Kl^{i_{n-k}}_{(1)n-k}\cdots l^{i_{n-2}}_{(1)n-2}\partial_{i_{n-k}}\cdots\partial_{i_{n-2}}\partial_j\rho^i{}_Ll^K_{(0)1}l^L_{(0)2}-{}\\
 &-l_{(0)1}\leftrightarrow l_{(0)2}-{}\\
 &-l^{i_{1}}_{1}\cdots l^{i_{n-k-2}}_{(1)n-k-2}\partial_{i_{1}}\cdots\partial_{i_{n-k-2}}T_{JKL}\eta^{IJ}l^{i_{n-k-1}}_{(1)n-k-1}\cdots l^{i_{n-2}}_{(1)n-2}\partial_{i_{n-k-1}}\cdots\partial_{i_{n-2}}\rho^i{}_Il^K_{(0)1}l^L_{(0)2}+{}\\
 &+\text{perm.}+\cdots
  \end{align*}
 \begin{align}
 &\Downarrow\nonumber\\
 \partial_{i_1}\cdots\partial_{i_{n-2}}(\rho^i{}_I\eta^{IJ}T_{JKL})&=\partial_{i_1}\cdots\partial_{i_{n-2}}(2\rho^j{}_{[K}\partial_{\underline{j}}\rho^i{}_{L]})\label{eq:cond21}
 \end{align}
 As above ``perm.'' indicates all possible unshuffles of $l_{(1)1},\ldots,l_{(1)n-2}$. Swapping $l_{(0)1}$ and $l_{(0)2}$ produces a sign since the Koszul sign is + but the parity of the permutation will be flipped, this introduces the antisymmetrisation in the final relation.\\
 \item\underline{$(l_{1},\ldots,l_n)=(l_{(1)1},\ldots,l_{(1)n-2},l_{(0)},l_{(-1)})$}
 \begin{align*}
 &\mu_1(\mu_n(l_{(1)1},\ldots,l_{(1)n-2},l_{(0)},l_{(-1)}))=\mu_2(\mu_{n-1}(l_{(1)1},\ldots,l_{(1)n-2},l_{(0)}),l_{(-1)})-{}\\
 &-\mu_2(\mu_{n-1}(l_{(1)1},\ldots,l_{(1)n-2},l_{(-1)}),l_{(0)}) -{}\\
 &-\mu_2(\mu_{n-1}(l_{(1)1},\ldots,l_{(1)n-3},l_{(0)},l_{(-1)}),l_{(1)n-2}) +\cdots+{}\\
 &+ \mu_{k+1}(\mu_{n-k}(l_{(1)1},\ldots,l_{(1)n-k-1},l_{(0)}),l_{(1)n-k},\ldots,l_{(1)n-2},l_{(-1)})+{}\\
 &+ (-1)^{k}\mu_{k+1}(\mu_{n-k}(l_{(1)1},\ldots,l_{(1)n-k-1},l_{(-1)}),l_{(1)n-k},\ldots,l_{(1)n-2},l_{(0)})-{}\\
 &- \mu_{k+1}(\mu_{n-k}(l_{(1)1},\ldots,l_{(1)n-k-2},l_{(0)},l_{(-1)}),l_{(1)n-k-1},\ldots,l_{(1)n-2})+{}\\
 &+\text{perm.}+\cdots\\
 &-l^{i_{1}}_{(1)1}\cdots l^{i_{n-2}}_{(1)n-2}\partial_{i_{1}}\cdots\partial_{i_{n-2}}\partial_{i}\rho^j{}_K\rho^i{}_J\eta^{IJ}l_{(-1)j}l^K_{(0)}={}\\
 &= \cdots+{}\\
 &+l_{(-1)j}l^K_{(0)}(-l^{i_{1}}_{(1)1}\cdots l^{i_{n-k-1}}_{(1)n-k-1}\partial_{i_{1}}\cdots\partial_{i_{n-k-1}}\rho^i{}_Kl^{i_{n-k}}_{(1)n-k}\cdots l^{i_{n-2}}_{(1)n-2}\partial_{i}\partial_{i_{n-k}}\cdots\partial_{i_{n-2}}\rho^j{}_J+{}\\
 &+l^{i_{1}}_{(1)1}\cdots l^{i_{n-k-1}}_{(1)n-k-1}\partial_{i_{1}}\cdots\partial_{i_{n-k-1}}\rho^j{}_L\eta^{LM}l^{i_{n-k}}_{(1)n-k}\cdots l^{i_{n-2}}_{(1)n-2}\partial_{i_{n-k}}\cdots\partial_{i_{n-2}}T_{JMK}\eta^{IJ}+{}\\
 &+l^{i_{1}}_{(1)1}\cdots l^{i_{n-k-2}}_{(1)n-k-2}\partial_{i_{1}}\cdots\partial_{i_{n-k-2}}\partial_{i}\rho^j{}_Kl^{i_{n-k-1}}_{(1)n-k-1}\cdots l^{i_{n-2}}_{(1)n-2}\partial_{i}\partial_{i_{n-k-1}}\cdots\partial_{i_{n-2}}\rho^i{}_J\eta^{IJ})+{}\\
 &+\text{perm.}+\cdots
  \end{align*}
  \begin{align}
 &\Downarrow\nonumber\\
 \partial_{i_1}\cdots\partial_{i_{n-2}}(\rho^j{}_L\eta^{LM}T_{MJK})&=\partial_{i_1}\cdots\partial_{i_{n-2}}(2\rho^i{}_{[J}\partial_{\underline{i}}\rho^j{}_{K]})\label{eq:cond22}
 \end{align}
 As above.\\
 \item\underline{$(l_{1},\ldots,l_n)=(l_{(1)1},\ldots,l_{(1)n-3},l_{(0)1},l_{(0)2},l_{(0)3})$}
 \begin{align*}
 &\mu_1(\mu_n(l_{(1)1},\ldots,l_{(1)n-3},l_{(0)1},l_{(0)2},l_{(0)3}))={}\\
 &=(-1)^{k+1}\mu_{k+1}(\mu_{n-k}(l_{(1)1},\ldots,l_{(1)n-k-2},l_{(0)1},l_{(0)2}),l_{(1)n-k-1},\ldots,l_{(1)n-3},l_{(0)3})-{}\\
 &-\mu_{k+1}(\mu_{n-k}(l_{(1)1},\ldots,l_{(1)n-k-1},l_{(0)1}),l_{(1)n-k},\ldots,l_{(1)n-3},l_{(0)2},l_{(0)3})-{}\\
 &-\mu_{k+1}(\mu_{n-k}(l_{(1)1},\ldots,l_{(1)n-k-3},l_{(0)1},l_{(0)2},l_{(0)3}),l_{(1)n-k-2},\ldots,l_{(1)n-3})+{}\\
 &+\text{perm.}+\cdots\\
 &-l^{i_{1}}_{(1)1}\cdots l^{i_{n-3}}_{(1)n-3}\partial_{i_{1}}\cdots\partial_{i_{n-3}}\partial_{i}\T_{ABC}\rho^i{}_J\eta^{IJ}l^A_{(0)1}l^B_{(0)2}l^C_{(0)3}={}\\
 &= \cdots+l^A_{(0)1}l^B_{(0)2}l^C_{(0)3}\cdot{}\\
 &\cdot(-l^{i_{1}}_{(1)1}\cdots l^{i_{n-k-1}}_{(1)n-k-1}\partial_{i_{1}}\cdots\partial_{i_{n-k-1}}\rho^i{}_Al^{i_{n-k}}_{(1)n-k}\cdots l^{i_{n-3}}_{(1)n-3}\partial_{i}\partial_{i_{n-k}}\cdots\partial_{i_{n-3}}T_{JBC}+{}\\
 &+l^{i_{1}}_{(1)1}\cdots l^{i_{n-k-2}}_{(1)n-k-2}\partial_{i_{1}}\cdots\partial_{i_{n-k-2}}T_{KAB}\eta^{KL}l^{i_{n-k-1}}_{(1)n-k-1}\cdots l^{i_{n-3}}_{(1)n-3}\partial_{i_{n-k-1}}\cdots\partial_{i_{n-3}}T_{JLC}\eta^{IJ}+{}\\
 &+l^{i_{1}}_{(1)1}\cdots l^{i_{n-k-3}}_{(1)n-k-3}\partial_{i_{1}}\cdots\partial_{i_{n-k-3}}\partial_{i}T_{ABC}l^{i_{n-k-2}}_{(1)n-k-2}\cdots l^{i_{n-3}}_{(1)n-3}\partial_{i_{n-k-2}}\cdots\partial_{i_{n-3}}\rho^i{}_J\eta^{IJ})+{}\\
 &+\text{perm.}+\cdots\\
  \end{align*}
 \begin{align}
 &\Downarrow\nonumber\\
 \partial_{i_1}\cdots\partial_{i_{n-3}}(\rho^i{}_J\partial_{i}T_{ABC}-3\rho^i{}_{[A}\partial_{\underline{i}}T_{BC]J}+3T_{JK[A}\eta^{KL}T_{BC]L})&=0\label{eq:cond31}
 \end{align}
 For reasons above, the graded Koszul sign induces the antisymmetrisation of $l_{(0)1},l_{(0)2},l_{(0)3}$.\\
 \item\underline{$(l_{1},\ldots,l_n)=(l_{(1)1},\ldots,l_{(1)n-3},l_{(0)1},l_{(0)2},l_{(-1)})$}
 \begin{align*}
 &0=-\mu_{k+1}(\mu_{n-k}(l_{(1)1},\ldots,l_{(1)n-k-1},l_{(0)1}),l_{(1)n-k},\ldots,l_{(1)n-3},l_{(0)2},l_{(-1)})+{}\\
 &+(-1)^{k+1}\mu_{k+1}(\mu_{n-k}(l_{(1)1},\ldots,l_{(1)n-k-1},l_{(-1)}),l_{(1)n-k},\ldots,l_{(1)n-3},l_{(0)1},l_{(0)2})+{}\\
 &+(-1)^{k+1}\mu_{k+1}(\mu_{n-k}(l_{(1)1},\ldots,l_{(1)n-k-2},l_{(0)1},l_{(0)2}),l_{(1)n-k-1},\ldots,l_{(1)n-3},l_{(-1)})-{}\\
 &-\mu_{k+1}(\mu_{n-k}(l_{(1)1},\ldots,l_{(1)n-k-2},l_{(0)1},l_{(-1)}),l_{(1)n-k-1},\ldots,l_{(1)n-3},l_{(0)2})+{}\\
 &+\text{perm.}+\cdots\\
 &0= \cdots+l^K_{(0)1}l^L_{(0)2}l_{(-1)j}\cdot{}\\
 &\cdot(-l^{i_{1}}_{(1)1}\cdots l^{i_{n-k-1}}_{(1)n-k-1}\partial_{i_{1}}\cdots\partial_{i_{n-k-1}}\rho^k{}_Kl^{i_{n-k}}_{(1)n-k}\cdots l^{i_{n-3}}_{(1)n-3}\partial_{k}\partial_{i_{n-k}}\cdots\partial_{i_{n-3}}\partial_{i}\rho^j{}_L+{}\\
 &+l^{i_{1}}_{(1)1}\cdots l^{i_{n-k-1}}_{(1)n-k-1}\partial_{i_{1}}\cdots\partial_{i_{n-k-1}}\rho^j{}_I\eta^{IJ}l^{i_{n-k}}_{(1)n-k}\cdots l^{i_{n-3}}_{(1)n-3}\partial_{i_{n-k}}\cdots\partial_{i_{n-3}}\partial_{i}T_{JKL}+{}\\
 &+l^{i_{1}}_{(1)1}\cdots l^{i_{n-k-2}}_{(1)n-k-2}\partial_{i_{1}}\cdots\partial_{i_{n-k-2}}T_{JKL}\eta^{IJ}l^{i_{n-k-1}}_{(1)n-k-1}\cdots l^{i_{n-3}}_{(1)n-3}\partial_{i_{n-k-1}}\cdots\partial_{i_{n-3}}\rho^j{}_I+{}\\
 &+l^{i_{1}}_{(1)1}\cdots l^{i_{n-k-2}}_{(1)n-k-2}\partial_{i_{1}}\cdots\partial_{i_{n-k-2}}\partial_{k}\rho^j{}_Kl^{i_{n-k-1}}_{(1)n-k-1}\cdots l^{i_{n-3}}_{(1)n-3}\partial_{i_{n-k-1}}\cdots\partial_{i_{n-3}}\partial_{i}\rho^k{}_L)+{}\\
 &+\text{perm.}+\cdots
  \end{align*}
 \begin{align}
 &\Downarrow\nonumber\\
 \partial_{i_1}\cdots\partial_{i_{n-3}}\partial_{i}(\rho^j{}_I\eta^{IJ}T_{JKL}-2\rho^k{}_{[K}\partial_{\underline{k}}\rho^j{}_{L]})&=0\label{eq:cond23}
 \end{align}
 As above.\\
 \item \underline{$(l_{1},\ldots,l_n)=(l_{(1)1},\ldots,l_{(1)n-4},l_{(0)1},l_{(0)2},l_{(0)3},l_{(0)4})$}
  \begin{align*}
 &0=\cdots+\mu_{k+1}(\mu_{n-k}(l_{(1)1},\ldots,l_{(1)n-k-1},l_{(0)1}),l_{(1)n-k},\ldots,l_{(1)n-4},l_{(0)2},l_{(0)3},l_{(0)4})+{}\\
 &+(-1)^{k+1}\mu_{k+1}(\mu_{n-k}(l_{(1)1},\ldots,l_{(1)n-k-2},l_{(0)1},l_{(0)2}),l_{(1)n-k-1},\ldots,l_{(1)n-4},l_{(0)3},l_{(0)4})+{}\\
 &+\mu_{k+1}(\mu_{n-k}(l_{(1)1},\ldots,l_{(1)n-k-3},l_{(0)1},l_{(0)2},l_{(0)3}),l_{(1)n-k-2},\ldots,l_{(1)n-4},l_{(0)4})+{}\\
 &+\text{perm.}+\cdots\\
 &0= \cdots+l^A_{(0)1}l^B_{(0)2}l^C_{(0)3}l^D_{(0)4}\cdot{}\\
 &\cdot(l^{i_{1}}_{(1)1}\cdots l^{i_{n-k-1}}_{(1)n-k-1}\partial_{i_{1}}\cdots\partial_{i_{n-k-1}}\rho^i{}_Al^{i_{n-k}}_{(1)n-k}\cdots l^{i_{n-4}}_{(1)n-4}\partial_{j}\partial_{i_{n-k}}\cdots\partial_{i_{n-4}}\partial_{i}T_{BCD}-{}\\
 &-l^{i_{1}}_{(1)1}\cdots l^{i_{n-k-2}}_{(1)n-k-2}\partial_{i_{1}}\cdots\partial_{i_{n-k-2}}T_{JAB}\eta^{IJ}l^{i_{n-k-1}}_{(1)n-k-1}\cdots l^{i_{n-4}}_{(1)n-4}\partial_{i_{n-k-1}}\cdots\partial_{i_{n-4}}\partial_{i}T_{ICD}-{}\\
 &-l^{i_{1}}_{(1)1}\cdots l^{i_{n-k-3}}_{(1)n-k-3}\partial_{i_{1}}\cdots\partial_{i_{n-k-3}}\partial_{j}T_{ABC}l^{i_{n-k-2}}_{(1)n-k-2}\cdots l^{i_{n-4}}_{(1)n-4}\partial_{i_{n-k-2}}\cdots\partial_{i_{n-4}}\partial_{i}\rho^j{}_D)+{}\\
 &+\text{perm.}+\cdots
  \end{align*}
 \begin{align}
 &\Downarrow\nonumber\\
 \partial_{i_1}\cdots\partial_{i_{n-4}}\partial_{i}(4\rho^j{}_{[A}\partial_{\underline{j}}T_{BCD]}-3T_{J[AB}\eta^{IJ}T_{CD]I})&=0\label{eq:cond32}
 \end{align}
 Equivalently as above the $\mathsf{L}$-degree of $l_{(0)i}$ and the antisymmetry of $T$ ensures the total antisymmetry in indices $A, B, C$ and $D$.
\end{itemize}
 It is immediately obvious that relations \eqref{eq:cond11} and \eqref{eq:cond12} are equivalent, as are \eqref{eq:cond21}, \eqref{eq:cond22} and \eqref{eq:cond23}, and that \eqref{eq:cond31} implies \eqref{eq:cond32}. Therefore, we have three unique sets of conditions giving all terms in the Taylor expansions of the axioms of the Courant algebroid  by taking $l_{(1)i}=X_i$:
 \begin{align*}
 \eta^{IJ}\rho^i{}_I(X)\rho^j{}_J(X)&=0,\\
 2\rho^j{}_{[I}(X)\partial_{\underline{j}}\rho^i{}_{J]}(X)-\rho^i{}_M(X)\eta^{ML}T_{LIJ}(X)&=0,\\
 3\rho^i{}_{[A}(X)\partial_{\underline{i}}T_{BC]J}(X)-\rho^i{}_J(X)\partial_{i}T_{ABC}(X)-3T_{JK[A}(X)\eta^{KM}T_{BC]M}(X)&=0.
 \end{align*}

\section{BV-BRST action for Courant sigma model}\label{app:BVappendix}

For completeness we write explicitly the action and BRST transformations for all fields in BV-BRST action for Courant sigma model.
\begin{align}
\begin{split}
S_\mathrm{BV}&=\int_{\Sigma_3}F  _i\dd{ X  }^i+\sfrac 1 2 \eta_{ {I} {J}} A  ^{ {I}} \dd{ A  }^{ {J}}
-\rho^i{}_{ {I}}( X  ) A  ^{ {I}} F  _i+\sfrac 1 6 T_{ {I} {J} {K}}( X  ) A  ^{ {I}} A  ^{ {J}} A  ^{ {K}}-{}\\
&\hphantom{\,=\,\int_{\Sigma_3}}-\epsilon^{  I}\rho^i{}_{  I}(X) X  ^\dagger_i+{}\\
&\hphantom{\,=\,\int_{\Sigma_3} }+\left(\dd{\epsilon}^{  I}+\eta^{ IJ}\rho^i{}_{  J}t_i+\eta^{ {I}{J} }T_{KLJ}(X)A  ^{  K}\epsilon^{ {L}} \right) A  ^\dagger_{ {I}}+{}\\
&\hphantom{\,=\,\int_{\Sigma_3} }+\left(\dd{t_i}-t_j \partial_i\rho^j{}_{  I}(X)A  ^{ {I}}-F  _j\partial_i\rho^j{}_{ {I}}(X)\epsilon^{  I} +\sfrac 1 2 \partial_iT_{ {I} {J} {K}}(X)A  ^{ I} A  ^{ J}\epsilon^{  K} \right) F  ^{\dagger i}+{}\\
&\hphantom{\,=\,\int_{\Sigma_3} }+\left(-\dd{v}_i- \partial_i\rho^j{}_{ {I}}(X)v_jA  ^{ {I}}+\partial_i\rho^j{}_{ {I}}(X)t_j\epsilon^{ {I}}+\sfrac 1 2 \partial_iT_{ {I} {J} {K}}(X)A  ^{  I}\epsilon^{  J}\epsilon^{  K} \right)t^{\dagger i}+{}\\
&\hphantom{\,=\,\int_{\Sigma_3} }+\left(-\eta^{ IJ}\rho^i{}_{ {J}}(X)v_i+\sfrac 1 2\eta^{  IJ} T_{ {J} {K} {L}}(X)\epsilon^{  K}\epsilon^{  L}\right)\epsilon^\dagger_{  I}+{}\\
&\hphantom{\,=\,\int_{\Sigma_3} }+\left(-\partial_i\rho^j{}_{  I}(X)v_j\epsilon^{ {I}} +\sfrac 1 6 \partial_iT_{ {I} {J} {K}}(X)\epsilon^{  I}\epsilon^{  J}\epsilon^{  K}\right)v^{\dagger i}+{}\\
&\hphantom{\,=\,\int_{\Sigma_3} }+
\left(\eta^{ IJ}\partial_i\rho^j{}_{ {J}}(X)v_j-\sfrac 1 2\eta^{  IJ} \partial_iT_{ {J} {K} {L}}(X)\epsilon^{  K}\epsilon^{  L}\right) F  ^{\dagger i} A  ^\dagger_{  I}+{}\\
&\hphantom{\,=\,\int_{\Sigma_3} }+\sfrac 1 2 \left( A  ^{ {I}}\partial_i\partial_j\rho^k{}_{ {I}}(X)v_k-t_k\partial_i\partial_j\rho^k{}_{ {I}}(X)\epsilon^{ {I}}-\sfrac 1 2 A  ^{  I} \partial_i\partial_jT_{ {I} {J} {K}}(X)\epsilon^{  J}\epsilon^{  K}\right) F  ^{\dagger i} F  ^{\dagger j}+{}\\
&\hphantom{\,=\,\int_{\Sigma_3} }+\left(-\partial_i\partial_j\rho^k{}_{  I}(X)v_k\epsilon^{ {I}}+\sfrac 1 6 \partial_i\partial_jT_{ {I} {J} {K}}(X)\epsilon^{  I}\epsilon^{  J}\epsilon^{  K}\right) F  ^{\dagger i}t^{\dagger j}+{}\\
&\hphantom{\,=\,\int_{\Sigma_3} }+\sfrac 1 6 \left(\partial_i\partial_j\partial_k\rho^l{}_{  I}(X)v_l\epsilon^{ {I}}-\sfrac 1 6 \partial_i\partial_j\partial_k T_{ {I} {J} {K}}(X)\epsilon^{  I}\epsilon^{  J}\epsilon^{  K}\right) F  ^{\dagger i} F  ^{\dagger j} F  ^{\dagger k},
\end{split}
\end{align}
and generalised BRST transformations for each field and antifield:
\begin{align}
Q_{\mathrm{BV}}X^i&=\rho^i{}_I(X)\epsilon^I,\label{eq:qbvx}\\
\begin{split}
Q_{\mathrm{BV}}A^I&=\dd\epsilon^I+\eta^{IJ}\rho^i{}_J(X)t_i+\eta^{IJ}T_{JKL}(X)A^K\epsilon^L+{}\\
&\hphantom{\,=\,}+F^{\dagger i}\eta^{IJ}\partial_{i}\rho^j{}_J(X)v_j-\sfrac 1 2 F^{\dagger i}\eta^{IJ}\partial_i T_{JKL}(X)\epsilon^K\epsilon^L,
\end{split}\\
\begin{split}
Q_{\mathrm{BV}}F_i&=\dd{t_i}-\partial_i\rho^j{}_{  J}(X)t_j A  ^{ {J}}-\partial_i\rho^j{}_{ {J}}(X)F  _j\epsilon^{  J} +\sfrac 1 2 \partial_iT_{ {I} {J} {K}}(X)A  ^{ I} A  ^{ J}\epsilon^{  K} -{}\\
&\hphantom{\,=\,}-\eta^{ IJ}\partial_i\rho^j{}_{ {J}}(X)v_jA^\dagger_I+\sfrac 1 2\eta^{  IJ} \partial_iT_{ {J} {K} {L}}(X)A^\dagger_I\epsilon^{  K}\epsilon^{  L}-t^{\dagger j}\partial_i\partial_j\rho^k{}_{  I}(X)v_k\epsilon^{ {I}}+{}\\
&\hphantom{\,=\,}+\sfrac 1 6 t^{\dagger j}\partial_i\partial_jT_{ {I} {J} {K}}(X)\epsilon^{  I}\epsilon^{  J}\epsilon^{  K}-F^{\dagger j}\partial_i\partial_j\rho^k{}_{ {I}}(X)v_kA  ^{ {I}}+F^{\dagger j}\partial_i\partial_j\rho^k{}_{ {I}}(X)t_k\epsilon^{ {I}}+{}\\
&\hphantom{\,=\,}+\sfrac 1 2 F^{\dag j} \partial_i\partial_jT_{ {I} {J} {K}}(X)A  ^{  I} \epsilon^{  J}\epsilon^{  K}+\sfrac 1 2 F  ^{\dagger j} F  ^{\dagger k}\partial_i\partial_j\partial_k\rho^l{}_{  I}(X)v_l\epsilon^{ {I}}-{}\\
&\hphantom{\,=\,}-\sfrac 1 {12} F  ^{\dagger j} F  ^{\dagger k}\partial_i\partial_j\partial_k T_{ {I} {J} {K}}(X)\epsilon^{  I}\epsilon^{  J}\epsilon^{  K},
\end{split}\\
Q_{\mathrm{BV}}\epsilon^I&=\eta^{ IJ}\rho^i{}_{ {J}}(X)v_i-\sfrac 1 2\eta^{  IJ} T_{ {J} {K} {L}}(X)\epsilon^{  K}\epsilon^{  L},\\
\begin{split}
Q_{\mathrm{BV}}t_i&=-\dd{v}_i- \partial_i\rho^j{}_{ {J}}(X)v_jA  ^{ {J}}+\partial_i\rho^j{}_{ {J}}(X)t_j\epsilon^{ {J}}+\sfrac 1 2 \partial_iT_{ {I} {J} {K}}(X)A  ^{  I}\epsilon^{  J}\epsilon^{  K}+\\
&\hphantom{\,=\,}+F^{\dagger j}\partial_i\partial_j\rho^k{}_{  J}(X)v_k\epsilon^{ {J}}-\sfrac 1 6 F^{\dagger j}\partial_i\partial_jT_{ {I} {J} {K}}(X)\epsilon^{  I}\epsilon^{  J}\epsilon^{  K},
\end{split}\\
Q_{\mathrm{BV}}v_i&=-\partial_i\rho^j{}_{  J}(X)v_j\epsilon^{ {J}} +\sfrac 1 6 \partial_iT_{ {I} {J} {K}}(X)\epsilon^{  I}\epsilon^{  J}\epsilon^{  K},\label{eq:qbvv}\\
\begin{split}\label{eq:qbvxdag}
Q_{\mathrm{BV}}X^\dagger_i&=-\dd F_i-\partial_{i}\rho^j{}_J(X)F_jA^J+\sfrac 1 6 \partial_{i}T_{IJK}(X)A^IA^JA^K+{}\\
&\hphantom{\,=\,}+\partial_{i}\rho^j{}_J(X)X^\dagger_j\epsilon^J-\partial_{i}\rho^j{}_J(X)v_j\epsilon^\dagger_K\eta^{JK}+\partial_{i}\rho^j{}_J(X)t_j\eta^{JK}A^\dagger_K-{}\\
&\hphantom{\,=\,}-\partial_{i}T_{IJK}(X)\eta^{IL}A^\dagger_L A^J\epsilon^K+\sfrac 1 2 \partial_{i}T_{IJK}(X)\epsilon^\dagger_L\eta^{IL}\epsilon^J\epsilon^K+{}\\
&\hphantom{\,=\,}+F^{\dagger j}\partial_{i}\partial_{j}\rho^k{}_I(X)v_k\eta^{IJ}A^\dagger_J+F^{\dagger j}\partial_{i}\partial_{j}\rho^k{}_I(X)t_kA^I+F^{\dagger j}\partial_{i}\partial_{j}\rho^k{}_I(X)F_k\epsilon^I-{}\\
&\hphantom{\,=\,}-\sfrac 1 2 F^{\dagger j}\partial_{i}\partial_{j}T_{IJK}(X)A^IA^J\epsilon^K-\sfrac 1 2 F^{\dagger j}\partial_{i}\partial_{j}T_{IJK}(X)\eta^{IL}A^\dagger_L\epsilon^J\epsilon^K-{}\\
&\hphantom{\,=\,}-t^{\dagger j}\partial_{i}\partial_{j}\rho^k{}_I(X)v_kA^I+t^{\dagger j}\partial_{i}\partial_{j}\rho^k{}_I(X)t_k\epsilon^I+\sfrac 1 2 t^{\dagger j}\partial_{i}\partial_{j}T_{IJK}(X)A^I\epsilon^J\epsilon^K+{}\\
&\hphantom{\,=\,}+v^{\dagger j}\partial_{i}\partial_{j}\rho^k{}_I(X)v_k\epsilon^I-\sfrac 1 6 v^{\dagger j}\partial_{i}\partial_{j}T_{IJK}(X)\epsilon^I\epsilon^J\epsilon^K+{}\\
&\hphantom{\,=\,}+\sfrac 1 2 F^{\dagger j}F^{\dagger k}\partial_{i}\partial_{j}\partial_{k}\rho^l{}_I(X)v_lA^I-\sfrac 1 2 F^{\dagger j}F^{\dagger k}\partial_{i}\partial_{j}\partial_{k}\rho^l{}_I(X)t_l\epsilon^I-{}\\
&\hphantom{\,=\,}-\sfrac 1 4 F^{\dagger j}F^{\dagger k}\partial_{i}\partial_{j}\partial_{k}T_{IJK}(X)A^I\epsilon^J\epsilon^K+F^{\dagger j}t^{\dagger k}\partial_{i}\partial_{j}\partial_{k}\rho^l{}_I(X)v_l\epsilon^I-{}\\
&\hphantom{\,=\,}-\sfrac 1 6 F^{\dagger j}t^{\dagger k}\partial_{i}\partial_{j}\partial_{k}T_{IJK}(X)\epsilon^I\epsilon^J\epsilon^K-\sfrac 1 6 F^{\dagger j}F^{\dagger k}F^{\dagger l}\partial_{i}\partial_{j}\partial_{k}\partial_{l}\rho^m{}_I(X)v_m\epsilon^I+{}\\
&\hphantom{\,=\,}+\sfrac 1 {36} F^{\dagger j}F^{\dagger k}F^{\dagger l}\partial_{i}\partial_{j}\partial_{k}\partial_{l}T_{IJK}(X)\epsilon^I\epsilon^J\epsilon^K,
\end{split}\\
\begin{split}
Q_{\mathrm{BV}}A^\dagger_I&=\eta_{IJ}\left(-\dd A^J+\eta^{JK}\rho^i{}_K(X)F_i-\sfrac 1 2 \eta^{JK}T_{KLM}(X)A^LA^M\right)-{}\\
&\hphantom{\,=\,}-F^{\dagger i}\partial_{i}\rho^j{}_I(X)t_j-F^{\dagger i}\partial_{i}T_{IJK}(X)A^J\epsilon^K-\sfrac 1 2 F^{\dagger i}F^{\dagger j}\partial_{i}\partial_{j}\rho^k{}_I(X)v_k+{}\\
&\hphantom{\,=\,}+\sfrac 1 4 F^{\dagger i}F^{\dagger j}\partial_{i}\partial_{j}T_{IJK}(X)\epsilon^J\epsilon^K+t^{\dagger i}\partial_{i}\rho^j{}_I(X)v_j-\sfrac 1 2 t^{\dagger i}\partial_{i}T_{IJK}(X)\epsilon^J\epsilon^K-{}\\
&\hphantom{\,=\,}- T_{IJK}(X)\eta^{JL}A^\dagger_L\epsilon^K,
\end{split}\\
\begin{split}\label{eq:qbvfdag}
Q_{\mathrm{BV}}F^{\dagger i}&=-\dd X^i+\rho^i{}_I(X)A^I-{}\\
&\hphantom{\,=\,}-F^{\dagger j}\partial_{j}\rho^i{}_I(X)\epsilon^I,
\end{split}\\
\begin{split}
Q_{\mathrm{BV}}\epsilon^\dagger_I&=\dd A^\dagger_I+\rho^i{}_I(X)X^\dagger_i-T_{IJK}(X)\eta^{JL}A^\dagger_LA^K+T_{IJK}(X)\eta^{JL}\epsilon^\dagger_L\epsilon^K+{}\\
&\hphantom{\,=\,}+F^{\dagger i}\partial_i\rho^j{}_I(X)F_j-\sfrac 1 2 F^{\dagger i}\partial_{i}T_{IJK}(X)A^JA^K- F^{\dagger i}\partial_{i}T_{IJK}(X)\eta^{JL}A^\dagger_L\epsilon^K+{}\\
&\hphantom{\,=\,}+t^{\dagger i}\partial_{i}\rho^j{}_I(X)t_j+t^{\dagger i}\partial_{i}T_{IJK}(X)A^J\epsilon^K+\sfrac 1 2 F^{\dagger i}F^{\dagger j}\partial_{i}\partial_{j}\rho^k{}_I(X)t_k-{}\\
&\hphantom{\,=\,}-\sfrac 1 2 F^{\dagger i}F^{\dagger j}\partial_{i}\partial_{j}T_{IJK}(X)A^J\epsilon^K+F^{\dagger i}t^{\dagger j}\partial_{i}\partial_{j}\rho^k{}_I(X)v_k-{}\\
&\hphantom{\,=\,}-\sfrac 1 2 F^{\dagger i}t^{\dagger j}\partial_{i}\partial_{j}T_{IJK}(X)\epsilon^J\epsilon^K-\sfrac 1 6 F^{\dagger i}F^{\dagger j}F^{\dagger k}\partial_{i}\partial_{j}\partial_{k}\rho^l{}_I(X)v_l+\\
&\hphantom{\,=\,}+\sfrac 1 {12} F^{\dagger i}F^{\dagger j}F^{\dagger k}\partial_{i}\partial_{j}\partial_{k}T_{IJK}(X)\epsilon^J\epsilon^K+v^{\dagger i}\partial_{i}\rho^j{}_I(X)v_j-{}\\
&\hphantom{\,=\,}-\sfrac 1 2 v^{\dagger i}\partial_{i}T_{IJK}(X)\epsilon^J\epsilon^K,
\end{split}\\
\begin{split}
Q_{\mathrm{BV}}t^{\dagger i}&=-\dd F^{\dagger i}+\eta^{IJ}\rho^i{}_I(X)A^\dagger_J+t^{\dagger j}\partial_{j}\rho^i{}_I(X)\epsilon^I +F^{\dagger j}\partial_{j}\rho^i{}_I(X)A^I-{}\\
&\hphantom{\,=\,}-\sfrac 1 2 F^{\dagger j}F^{\dagger k}\partial_{j}\partial_{k}\rho^i{}_I(X)\epsilon^I,
\end{split}\\
\begin{split}
Q_{\mathrm{BV}}v^{\dagger i}&=-\dd t^{\dagger i}+\eta^{IJ}\rho^i{}_I(X)\epsilon^\dagger_J-F^{\dagger j}\partial_{j}\rho^i{}_I(X)\eta^{IJ}A^\dagger_J+t^{\dagger j}\partial_{j}\rho^i{}_I(X)A^I-v^{\dagger i}\partial_{j}\rho^i{}_I(X)\epsilon^I-{}\\
&\hphantom{\,=\,}-\sfrac 1 2 F^{\dagger j}F^{\dagger k}\partial_{j}\partial_{k}\rho^i{}_I(X)A^I - F^{\dagger j}t^{\dagger k}\partial_{j}\partial_{k}\rho^i{}_I(X)\epsilon^I+\sfrac 1 6 F^{\dagger j}F^{\dagger k}F^{\dagger l}\partial_{j}\partial_{k}\partial_{l}\rho^i{}_I(X)\epsilon^I.
\end{split}
\end{align}
As was to be expected one may notice the classical part (first line in each expression) of the BRST transformations of physical and ghost fields \eqref{eq:qbvx}--\eqref{eq:qbvv} corresponds to their gauge variations \eqref{eq:xvar}--\eqref{eq:Fvar} and antifields \eqref{eq:qbvxdag}--\eqref{eq:qbvfdag} to their equations of motion \eqref{eq:Feom}--\eqref{eq:Xeom}.

\section{Homotopy identities of extended CA algebra}\label{sec:CAhomotopy}

Homotopy relations \eqref{eq:homotopyjac} imply the possible choices for the higher products i.e. restrict us in which can be set to vanish. In this section we will make the calculation for $\tilde{\mathsf{L}}_1=TM$ of which the restriction to $T_pM$, as in section \ref{subsec:Linftyextension}, is a special case. Therefore we have the following for $i=1,2,3,4$ and $i\geqslant4$.
\begin{itemize}
	\item \underline{$i=1$}\\
	There is only one non-trivial homotopy relation:
	\[\tilde{\mu}_1\tilde{\mu}_1(f)=\tilde\rho\circ\mathcal{D}(f)=0,\]
	which is satisfied by the axioms of the Courant algebroid.
	\item \underline{$i=2$}\\
	Of the four non-trivial relations, three will be modified by the existence of $\mathsf{L}_1$. Choices $(l_1,l_2)=(e,f),(h,f),(e_1,e_2)$ produce the following conditions respectively:
	\begin{align*}
	\tilde{\mu}_2(\tilde\rho(e),f)&=0,\\
	\tilde{\mu}_2(\mathcal{D}f,h)&=0,\\
	\tilde{\mu}_2(\tilde\rho(e_1),e_2)-\tilde{\mu}_2(\tilde\rho(e_2),e_1)&=[\tilde{\rho}(e_1),\tilde{\rho}(e_2)].\\
	\end{align*}
	The first enables us to set $\tilde{\mu}_2(h,f)=0$, whereas from the second and third relation it is obvious $\tilde{\mu}_2(h,e)$ cannot vanish and one can choose $\tilde{\mu}_2(h,e)^i=h^j\tilde{\partial}_j\tilde{\rho}(e)^i$.
	\item \underline{$i=3$}\\
	In this case there are six combinations of elements that produce non-trivial homotopy relations, of which five are modified by the extension: $(h,f_1,f_2),(h,e,f),(e_1,e_2,e_3),(h_1,h_2,f)$ and $(h,e_1,e_2)$. These combinations respectively produce the constraints:
	\begin{align*}
	\tilde{\mu}_3(\mathcal{D}f_1,h,f_2)+\tilde{\mu}_3(\mathcal{D}f_2,h,f_1)&=0,\\
	\tilde{\mu}_3(\tilde{\rho}(e),h,f)+\tilde{\mu}_3(\mathcal{D}f,h,f)&=0,\\
	\tilde{\mu}_3(\tilde{\rho}(e_1),e_2,e_3)+\text{cyclic}&=0,\\
	\tilde{\rho}(\tilde{\mu}_3(h_1,h_2,f))+\tilde{\mu}_3(\mathcal{D}f,h_1,h_2)&=0,\\
	\tilde{\mu}_3(\tilde{\rho}(e_1),e_2,h)^i+h^j\tilde{\rho}(e_1)^k\tilde{\partial}_j\tilde{\partial}_k\tilde{\rho}(e_2)^i-e_1\leftrightarrow e_2&=-\tilde{\rho}(\tilde{\mu}_3(h,e_1,e_2))^i.
	\end{align*}
	The minimal extension implied by these constraints is to set all $\tilde{\mu}_3$ products involving $h$ to zero except $\tilde{\mu}_3(h_1,h_2,e)^i=h_1^jh_2^k\tilde{\partial}_j\tilde{\partial}_k\tilde{\rho}(e)^i$.
	\item \underline{$i=4$}\\
	Degree counting tells us that for $i\geqslant4$ there are always exactly 7 non-trivial homotopy conditions, additionally, all of them are modified by the extension and must be calculated. The combinations are: $(h_1,h_2,f_1,f_2),(h,e_1,e_1,f),(e_1,e_2,e_3,e_4),(h_1,h_2,e,f),(h,e_1,e_2,e_3),(h_1,h_2,h_3,f)$ and $(h_1,h_2,e_1,e_2)$. In order, each combination produces the following conditions:
	\begin{align*}
	\tilde{\mu}_4(\mathcal{D}f_1,h_1,h_2,f_2)+\tilde{\mu}_4(\mathcal{D}f_2,h_1,h_2,f_1)&=0,\\
	\tilde{\mu}_4(\tilde{\rho}(e_1),h,e_2,f)-\tilde{\mu}_4(\tilde{\rho}(e_2),h,e_1,f)-\tilde{\mu}_4(\mathcal{D}f,h,e_1,e_2)&=0,\\
	\tilde{\mu}_4(\tilde{\rho}(e_{[1}),e_{2]},e_3,e_4)+\tilde{\mu}_4(\tilde{\rho}(e_{[3}),e_{4]},e_1,e_2)&=0,\\
	\mathcal{D}\tilde{\mu}_4(h_1,h_2,e,f)-\tilde{\mu}_4(\tilde{\rho}(e),h_1,h_2,f)+\tilde{\mu}_4(\mathcal{D}f,h_1,h_2,e)&=0,\\
	\tilde{\mu}_4(\tilde{\rho}(e_1),e_2,e_3,h)+\text{cyclic}&=\mathcal{D}\tilde{\mu}_4(e_1,e_2,e_3,h),\\
	\tilde{\rho}(\tilde{\mu}_4(h_1,h_2,h_3,f))-\tilde{\mu}_4(\mathcal{D}f,h_1,h_2,h_3)&=0,\\	
	\tilde{\mu}_4(\tilde{\rho}(e_1),e_2,h_1,h_2)^i+h_1^jh_2^k\tilde{\rho}(e_2)^l\tilde{\partial}_j\tilde{\partial}_k\tilde{\partial}_l\tilde{\rho}(e_1)^i-e_1\leftrightarrow e_2&=\tilde{\rho}(\tilde{\mu}_4(e_1,e_2,h_1,h_2))^i.
	\end{align*}
	These constraints allow the minimal extension of non-vanishing products to be just one: $\tilde{\mu}_4(h_1,h_2,h_3,e)^i=h_1^jh_2^kh_3^l\tilde{\partial}_j\tilde{\partial}_k\tilde{\partial}_l\tilde{\rho}(e)^i$.
	\item \underline{$i\geqslant4$}\\
	Since for $i$ greater than four there are no new types of homotopy relations, since the combinations of elements from $i=4$ that produce non-trivial homotopy identities all simply gain the appropriate number of $h$ elements. Therefore, the structures of the corresponding conditions placed upon $\tilde{\mu}_i$ will be no different from the case of $i=4$. For that reason we make the assumption that all higher products vanish except $\tilde{\mu}_i(h_1,\ldots,h_{i-1},e)$ since it could not be made to vanish in lower cases. To be consistent with our choice for $\tilde{\mu}_4$ we make the ansatz: \[\tilde{\mu}_{i}(h_1,\ldots,h_{i-1},e)^j=h_1^{j_1}\cdots h_{i-1}^{j_i}\tilde{\partial}_{j_1}\cdots\tilde{\partial}_{j_i}\tilde{\rho}(e)^j.\]
	The consequence of this assumption is that only two homotopy identities will be non-trivial, those corresponding to combinations: $(h_1,\ldots,h_{i-1},f)$ and $(h_1,\ldots,h_{i-2},e_1,e_2)$. The first is directly satisfied by the axioms of a Courant algebroid:
	\[\tilde{\mu}_i(h_1,\ldots,h_{i-1},\mathcal{D}f)=0.\]
	The second is just the higher derivative analogue to the final condition in the $i=4$ case:
	\begin{align*}
	0&=-\tilde{\mu}_2(\tilde{\mu}_{i-1}(h_1,\ldots,h_{i-2},e_1),e_2)+\tilde{\mu}_2(\tilde{\mu}_{i-1}(h_1,\ldots,h_{i-2},e_2),e_1)-{}\\
	&\phantom{\,=\,}-\cdots-{}\\
	&\phantom{\,=\,}-\tilde{\mu}_n(\tilde{\mu}_{i-n+1}(h_1,\ldots,h_{i-n},e_1),h_{i-n+1},\ldots,h_{i-2},e_2)-\text{perm.}+{}\\
	&\phantom{\,=\,}+\tilde{\mu}_n(\tilde{\mu}_{i-n+1}(h_1,\ldots,h_{i-n},e_2),h_{i-n+1},\ldots,h_{i-2},e_1)+\text{perm.}-{}\\
	&\phantom{\,=\,}-\cdots-{}\\
	&\phantom{\,=\,}-\tilde{\mu}_{i-1}(\tilde{\mu}_{2}(h_1,e_1),h_{2},\ldots,h_{i-2},e_2)-\text{perm.}+{}\\
	&\phantom{\,=\,}+\tilde{\mu}_{i-1}(\tilde{\mu}_{2}(h_1,e_2),h_{2},\ldots,h_{i-2},e_1)+\text{perm.}+{}\\
	&\phantom{\,=\,}+(-1)^{i}\tilde{\mu}_{i-1}(\tilde{\mu}_{2}(e_1,e_2),h_1,\ldots,h_{i-2})-{}\\
	&\phantom{\,=\,}-\tilde{\mu}_{i}(\tilde{\mu}_{1}(e_1),h_{1},\ldots,h_{i-2},e_2)+\tilde{\mu}_{i}(\tilde{\mu}_{1}(e_2),h_{1},\ldots,h_{i-2},e_1),\\
	\end{align*}
	which is directly satisfied by use of the ansatz and Leibniz rule of the differential operator $\tilde{\partial}_{i_1}\cdots\tilde{\partial}_{i_n}$.
\end{itemize}


\begin{thebibliography}{99}

\bibitem{csft}
  B.~Zwiebach,
 ``Closed string field theory: Quantum action and the B-V master equation,''
  Nucl.\ Phys.\ B {\bf 390} (1993) 33
\href{https://arxiv.org/abs/hep-th/9206084}{[{\tt arXiv:hep-th/9206084}]}.

\bibitem{ls}
  T.~Lada and J.~Stasheff,
  ``Introduction to SH Lie algebras for physicists,''
  Int.\ J.\ Theor.\ Phys.\  {\bf 32} (1993) 1087
  \href{https://arxiv.org/abs/hep-th/9209099}{[{\tt arXiv:hep-th/9209099}]}.
  
\bibitem{OB}
O.~Hohm and B.~Zwiebach,
``$L_{\infty}$ Algebras and Field Theory,''
Fortsch.\ Phys.\  {\bf 65} (2017) no.3-4,  1700014
\href{https://arxiv.org/abs/1701.08824}{[{\tt arXiv:1701.08824 [hep-th]}]}.


\bibitem{rb}
  R.~Blumenhagen, I.~Brunner, V.~Kupriyanov and D.~L\"{u}st,
``Bootstrapping non-commutative gauge theories from L$_\infty$ algebras,''
  JHEP {\bf 1805} (2018) 097
 \href{https://arxiv.org/abs/1803.00732}{[{\tt arXiv:1803.00732 [hep-th]}]}.
  
  \bibitem{glenn}
  G.~Barnich, F.~Brandt and M.~Henneaux,
 ``Local BRST cohomology in gauge theories,''
  Phys.\ Rept.\  {\bf 338} (2000) 439
  \href{https://arxiv.org/abs/hep-th/0002245}{[{\tt arXiv:hep-th/0002245}]}.
  
  \bibitem{S1}
W. Siegel, 
``Two vierbein formalism for string inspired axionic gravity,''
Phys. Rev. D {\bf 47} (1993) 5453
\href{https://arxiv.org/abs/hep-th/9302036}{[{\tt arXiv:hep-th/9302036}]}.

\bibitem{S2}
W. Siegel,
``Superspace duality in low-energy superstrings,'' 
 {Phys. Rev. D} {\bf 48} (1993) 2826 
\href{https://arxiv.org/abs/hep-th/9305073}{[{\tt arXiv:hep-th/9305073}]}.

\bibitem{HZ1}
C. Hull and B. Zwiebach,
``Double Field Theory,''
{JHEP} {\bf 09} (2009) 099 
\href{https://arxiv.org/abs/0904.4664}{[{\tt arXiv:0904.4664 [hep-th]}]}.



\bibitem{HZ2}
C. Hull and B. Zwiebach,
``The gauge algebra of double field theory and Courant brackets''
{JHEP} {\bf 09} (2009) 090
\href{https://arxiv.org/abs/0908.1792}{[{\tt arXiv:0908.1792 [hep-th]}]}.

\bibitem{C90}
T. J. Courant,
``Dirac manifolds,''
{Trans. Am. Math. Soc.} {\bf 319} (1990) 631.

\bibitem{LWX}
Z.-J. Liu, A. Weinstein and P. Xu,
``Manin Triples for Lie Bialgebroids, ''
J. Diff. Geom. {\bf 45} (1997) 547
\href{https://arxiv.org/abs/dg-ga/9508013}{[{\tt arXiv:dg-ga/9508013}]}.


\bibitem{Sev}
P. \v Severa,
``Letters to Alan Weinstein about Courant algebroids, ''
\href{https://arxiv.org/abs/1707.00265}{{\tt arXiv:1707.00265 [math.DG]}}.

\bibitem{Roytenberg:1998vn}
D.~Roytenberg and A.~Weinstein,
``Courant Algebroids and Strongly Homotopy Lie Algebras,'' 
Lett. Math. Phys. {\bf 46} (1998) 81
\href{https://arxiv.org/abs/math/9802118}{[{\tt arXiv:math/9802118 [math.QA]}]}.

\bibitem{Roytenberg:2006qz}
D.~Roytenberg,
``AKSZ-BV Formalism and Courant Algebroid-induced Topological Field Theories,''
Lett.\ Math.\ Phys.\  {\bf 79} (2007) 143
\href{https://arxiv.org/abs/hep-th/0608150}{[{\tt arXiv:hep-th/0608150}]}.
%

\bibitem{P1}
J.-S.~Park, ``Topological open p-branes,” in: Symplectic Geometry and Mirror Symmetry, eds.
K. Fukaya, Y.-G. Oh, K. Ono and G. Tian (World Scientific, 2001), pp. 311–384
\href{https://arxiv.org/abs/hep-th/0012141}{[{\tt arXiv:hep-th/0012141}]}.

\bibitem{Ikeda}
N.~Ikeda,
``Chern-Simons gauge theory coupled with BF theory,''
Int.\ J.\ Mod.\ Phys.\ A {\bf 18} (2003) 2689
\href{https://arxiv.org/abs/hep-th/0203043}{[{\tt arXiv:hep-th/0203043}]}.

\bibitem{HP}
C.~Hofman and J.-S.~Park, ``BV quantization of topological open membranes,” Commun. \ Math. \  Phys.
{\bf 249} (2004) 249  \href{https://arxiv.org/abs/hep-th/0209214}{[{\tt arXiv:hep-th/0209214}]}.

\bibitem{AKSZ}
  M.~Alexandrov, M.~Kontsevich,A.~Schwarz, and  O.~Zaboronsky 
 ``The Geometry of the master equation and topological quantum field theory,''
  Int.\ J.\ Mod.\ Phys.\ A {\bf 12} (1997) 1405
    \href{https://arxiv.org/abs/hep-th/9502010}{[{\tt arXiv:hep-th/9502010}]}.

\bibitem{Hal}
N. Halmagyi,
``Non-geometric backgrounds and the first order string sigma-model,''
\href{https://arxiv.org/abs/0906.2891}{{\tt arXiv:0906.2891 [hep-th]}}.

 \bibitem{Mylonas}
  D.~Mylonas, P.~Schupp and R.~J.~Szabo,
  ``Membrane Sigma-Models and Quantization of Non-Geometric Flux Backgrounds,''
  JHEP {\bf 1209} (2012) 012
  \href{https://arxiv.org/abs/1207.0926}{[{\tt arXiv:1207.0926 [hep-th]}]}.
  
  \bibitem{ChJL}
  A.~Chatzistavrakidis, L.~Jonke and O.~Lechtenfeld,
``Sigma models for genuinely non-geometric backgrounds,''
  JHEP {\bf 1511} (2015) 182
  \href{https://arxiv.org/abs/1505.05457}{[{\tt arXiv:1505.05457 [hep-th]}]}.
  
  \bibitem{Watamura}
  T.~Bessho, M.~A.~Heller, N.~Ikeda and S.~Watamura,
``Topological Membranes, Current Algebras and H-flux - R-flux Duality based on Courant Algebroids,''
  JHEP {\bf 1604} (2016) 170
  \href{https://arxiv.org/abs/1511.03425}{[{\tt arXiv:1511.03425 [hep-th]}]}.


\bibitem{p12}
A. Chatzistavrakidis, L. Jonke, F. S. Khoo and R. J. Szabo,
``Double Field Theory and Membrane Sigma-Models,''
{JHEP} {\bf 1807} (2018) 015
\href{https://arxiv.org/abs/1802.07003}{[{\tt arXiv:1802.07003 [hep-th]}]}.


\bibitem{brano} 
B.~Jur\v{c}o, L.~Raspollini, C.~S\"{a}mann and M.~Wolf,
``$L_\infty$-Algebras of Classical Field Theories and the Batalin-Vilkovisky Formalism,''
Fortsch.\ Phys.\  {\bf 67}, no. 7, 1900025 (2019)
\href{https://arxiv.org/abs/1809.09899}{[{\tt arXiv:1809.09899 [hep-th]}]}.

\bibitem{gs}
M.~Gr\"{u}tzmann and T.~Strobl,
  ``General Yang-Mills type gauge theories for $p$-form gauge fields: From physics-based ideas to a mathematical framework or From Bianchi identities to twisted Courant algebroids,''
  Int.\ J.\ Geom.\ Meth.\ Mod.\ Phys.\  {\bf 12} (2014) 1550009
  \href{https://arxiv.org/abs/1407.6759}{[{\tt arXiv:1407.6759 [hep-th]}]}.
  
\bibitem{glob}
C.~C\'ordova, T.~T.~Dumitrescu and K.~Intriligator,
  ``Exploring 2-Group Global Symmetries,''
  JHEP {\bf 1902} (2019) 184
  \href{https://arxiv.org/abs/1802.04790}{[{\tt arXiv:1802.04790 [hep-th]}]}.
  
 
  \bibitem{proc}
  A.~Chatzistavrakidis, C.~J.~Grewcoe, L.~Jonke, F.~S.~Khoo and R.~J.~Szabo,
  ``BRST symmetry of doubled membrane sigma-models,''
  PoS CORFU {\bf 2018} (2019) 147
  \href{https://arxiv.org/abs/1904.04857}{[{\tt arXiv:1904.04857 [hep-th]}]}.
  

\bibitem{Voronov}
Th.~Voronov, 
  ``Higher derived brackets and homotopy algebras,'' 
J. Pure and Appl. Algebra. {\bf 202} (2005)  133 \href{https://arxiv.org/abs/math/0304038}{[{\tt arXiv:math/0304038 [math.QA]}]}.

\bibitem{dee}
	D.~Roytenberg, 
	``On the structure of graded symplectic supermanifolds and
        Courant algebroids,''
        Contemp. Math. {\bf 315} (2002) 169--186
	\href{https://arxiv.org/abs/math/0203110}{[{\tt arXiv:math/0203110 [math.SG]}]}.

\bibitem{fulp}
R.~Fulp, T.~Lada and J.~Stasheff,
  ``Noether's variational theorem II and the BV formalism,''
  Rend.\ Circ.\ Mat.\ Palermo S {\bf 71} (2003) 115
 \href{https://arxiv.org/abs/math/0204079}{[{\tt arxiv:math/0204079 [math-qa]}]}.

\end{thebibliography}
\end{document}